\documentclass[nofootinbib,10pt,twocolumn,preprintnumbers]{revtex4-1}
\usepackage{amsmath,amssymb,pgf,pgfarrows,pgfnodes,float,appendix, hyperref,slashed,breakurl,graphicx}
\definecolor{Color}{rgb}{0.28, 0.24, 0.55}
\definecolor{Orange}{rgb}{1,0.38,0.11}
\hypersetup{
    colorlinks = true,
    citecolor = Orange,
    linkcolor = Color,
    urlcolor  = Orange,
}
\definecolor{internationalorange}{rgb}{1.0, 0.31, 0.0}

\usepackage{graphicx}
\usepackage{subfigure}
\usepackage[margin=0.9in]{geometry}

\usepackage{enumerate}

\usepackage{amsmath}

\usepackage{color, colortbl}
\definecolor{Gray}{gray}{0.8}
\definecolor{GrayLight}{gray}{0.4}
\definecolor{Darkgreen}{RGB}{30,120,30}
\definecolor{granate}{rgb}{0.8039,0.2,0.2}
\newcommand{\beq}{\begin{equation}}
\newcommand{\eeq}{\end{equation}}
\newcommand{\bea}{\begin{eqnarray}}
\newcommand{\eea}{\end{eqnarray}}

\usepackage{eso-pic}

\usepackage{xparse}

\usepackage{tikz}
\usetikzlibrary{"arrows", "automata", "backgrounds", "calendar", "chains", "matrix", "mindmap", "patterns", "petri", "shadows", "shapes.geometric", "shapes.misc", "spy", "trees"}
\usetikzlibrary{arrows,shapes}
\usetikzlibrary{trees}
\usetikzlibrary{matrix,arrows} 				
\usetikzlibrary{positioning}				
\usetikzlibrary{calc,through}				
\usetikzlibrary{decorations.pathreplacing}  
\usepackage{pgffor}							

\usetikzlibrary{decorations.pathmorphing}	
\usetikzlibrary{decorations.markings}
\tikzset{
    vector/.style={decorate, decoration={snake}, draw},
	provector/.style={decorate, decoration={snake,amplitude=2.5pt}, draw},
	antivector/.style={decorate, decoration={snake,amplitude=-2.5pt}, draw},
    fermion/.style={draw=black, postaction={decorate},
        decoration={markings,mark=at position .55 with {\arrow[draw=black]{>}}}},
    fermionr/.style={draw=black, postaction={decorate},
    decoration={markings,mark=at position .55 with {\arrow[draw=black]{<}}}},
    fermioncyan/.style={draw=black, postaction={decorate},
        decoration={markings,mark=at position .55 with {\arrow[draw=cyan]{<}}}},
    fermiondif/.style={draw=black, postaction={decorate},
        decoration={markings,mark=at position .7 with {\arrow[draw=black]{>}}}},
            fermiondif2/.style={draw=black, postaction={decorate},
        decoration={markings,mark=at position .7 with {\arrow[draw=black]{<}}}},
    fermionend/.style={draw=black, postaction={decorate},
        decoration={markings,mark=at position 1 with {\arrow[draw=black]{>}}}},
    fermionuchannel2/.style={draw=black, postaction={decorate},
        decoration={markings,mark=at position .4 with {\arrow[draw=black]{>}}}},
    scalardif/.style={dashed,draw=black, postaction={decorate},
        decoration={markings,mark=at position .7 with {\arrow[draw=black]{>}}}},
    scalarend/.style={dashed,draw=black, postaction={decorate},
        decoration={markings,mark=at position 1 with {\arrow[draw=black]{>}}}},
    fermionbar/.style={draw=black, postaction={decorate},
        decoration={markings,mark=at position .55 with {\arrow[draw=black]{<}}}},
    fermionnoarrow/.style={draw=black},
    gluon/.style={decorate, draw=black,
        decoration={coil,amplitude=4pt, segment length=5pt}},
    scalar/.style={dashed,draw=black, postaction={decorate},
        decoration={markings,mark=at position .55 with {\arrow[draw=black]{>}}}},
    scalarcyan/.style={dashed,draw=black, postaction={decorate},
        decoration={markings,mark=at position .55 with {\arrow[draw=cyan]{>}}}},
    scalaruchannel1/.style={dashed,draw=black, postaction={decorate},
        decoration={markings,mark=at position .7 with {\arrow[draw=black]{>}}}},
                  scalaruchannel2/.style={dashed,draw=black, postaction={decorate},
        decoration={markings,mark=at position .4 with {\arrow[draw=black]{>}}}},
    scalarbar/.style={dashed,draw=black, postaction={decorate},
        decoration={markings,mark=at position .55 with {\arrow[draw=black]{<}}}},
    scalarnoarrow/.style={dashed,draw=black},
    electron/.style={draw=black, postaction={decorate},
        decoration={markings,mark=at position .55 with {\arrow[draw=black]{>}}}},
	bigvector/.style={decorate, decoration={snake,amplitude=4pt}, draw},
}

\NewDocumentCommand\semiloop{O{black}mmmO{}O{above}}
{%
\draw[#1] let \p1 = ($(#3)-(#2)$) in (#3) arc (#4:({#4+180}):({0.5*veclen(\x1,\y1)})node[midway, #6] {#5};)
}

\tikzstyle{block} = [draw, rectangle, 
    minimum height=3em, minimum width=6em]

\usepackage{enumitem}
\usepackage{tikz}
\usetikzlibrary{fit}
\tikzset{%
  highlight/.style={rectangle,rounded corners,color=granate,draw,text opacity =1,
    fill opacity=0.5,thick,inner sep=0pt}
}

\usepackage{xparse}
\NewDocumentCommand\loopv{O{black}mmmO{}O{above}}
{%
\draw[#1] let \p1 = ($(#3)-(#2)$) in (#3) arc (#4:({#4+360}):({0.5*veclen(\x1,\y1)})node[midway, #6] {#5};)
}

\usepackage{placeins}

\tikzset{
    cross/.pic = {
    \draw[rotate = 45] (-#1,0) -- (#1,0);
    \draw[rotate = 45] (0,-#1) -- (0, #1);
    }
}

\tikzset{
    square/.style={%
        draw=none,
        circle,
        append after command={%
            \pgfextra \draw[#1] (\tikzlastnode.north-|\tikzlastnode.west) rectangle 
                (\tikzlastnode.south-|\tikzlastnode.east);\endpgfextra}
    },
    square/.default=black
}

\tikzstyle{block} = [draw, rectangle, 
    minimum height=3em, minimum width=6em]

\usepackage{xparse}

\begin{document}

\title{\Large{Majorana Neutrinos and Dark Matter from Anomaly Cancellation}}
\author{Hridoy Debnath, Pavel Fileviez P\'erez, Kevin Gonz\'alez-Quesada}
\affiliation{
Physics Department and Center for Education and Research in Cosmology and Astrophysics (CERCA), Case Western Reserve University, Cleveland, OH 44106, USA}
\email{hxd253@case.edu, pxf112@case.edu, kag155@case.edu}

\begin{abstract}
We discuss a simple theory for neutrino masses where the total lepton number is a local gauge symmetry spontaneously broken below the multi-TeV scale. In this context, the neutrino masses are generated through the canonical seesaw mechanism and a Majorana dark matter candidate is predicted from anomaly cancellation. We discuss in great detail the dark matter annihilation channels and find out the upper bound on the symmetry breaking scale using the cosmological bounds on the relic density. Since in this context the dark matter candidate has suppressed couplings to the Standard Model quarks, one can satisfy the direct detection bounds even if the dark matter mass is close to the electroweak scale. This theory predicts a light pseudo-Nambu-Goldstone boson (the Majoron) associated to the mechanism of neutrino mass. We discuss briefly the properties of the Majoron and the impact of the Big Bang Nucleosynthesis bounds.
\end{abstract}

\maketitle

\section{INTRODUCTION}
The origin of neutrino masses is one of the most pressing issues in particle physics. It is well-known that the Standard Model of Particle Physics is one of the most successful theories of nature but it does not provide a mechanism to understand the origin of neutrino masses. The Standard Model neutrinos could be Dirac or Majorana fermions. The most popular idea to explain the smallness of Majorana neutrino masses is based on the seesaw mechanism~\cite{Minkowski:1977sc,Gell-Mann:1979vob,Yanagida:1979as,Mohapatra:1979ia}. 

In the context of the canonical seesaw mechanism, the neutrino masses are suppressed by the mass of the hypothetical heavy right-handed neutrinos. Unfortunately, the relevant scale for the seesaw mechanism can be as large as $10^{14} - 10^{15}$ GeV. We could test directly the origin of neutrino masses if the mechanism is realized at energy scales that one can reach at colliders or other experiments. In this article, we discuss a simple theory where the seesaw scale is below the multi-TeV scale and one can hope to test the origin of neutrino masses in the near future.
One can consider a simple theory for Majorana neutrino masses based on the spontaneous breaking of the total lepton number~\cite{FileviezPerez:2011pt,Duerr:2013dza,FileviezPerez:2014lnj}. In this theory, the total lepton number is a local gauge symmetry. Since the total lepton number is not an anomaly-free symmetry in the Standard Model, one needs to add extra fermionic fields to cancel all anomalies and study the spontaneous breaking of this symmetry. Different solutions have been proposed to this issue in Refs.\cite{FileviezPerez:2011pt,Duerr:2013dza,FileviezPerez:2014lnj}, see also Ref.~\cite{Foot:1989ts} for an earlier discussion. In most of these solutions one predicts a fermionic dark matter candidate from anomaly cancellation. Therefore, a simple theory for Majorana neutrino masses predicts a candidate for the cold dark matter in the universe. See Fig.~\ref{idea} for the main idea that defines the connection between the origin of neutrino masses, anomaly cancellation and the existence of a fermionic dark matter candidate. 
\begin{widetext}
\begin{center}
\begin{figure}[h]
\centering
\includegraphics[width=0.6\textwidth]{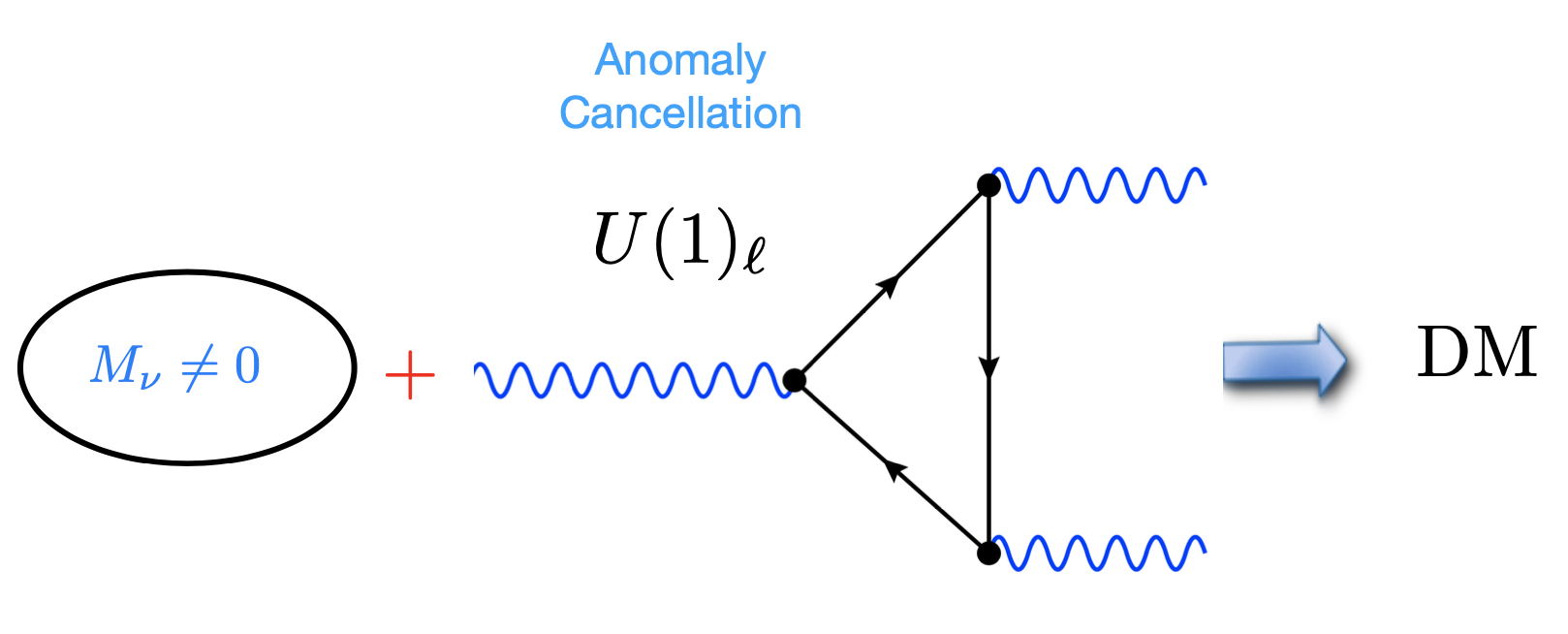}
\caption{Relation between the origin of neutrino masses, anomaly cancellation and dark matter.}
\label{idea}
\end{figure}
\end{center}
\end{widetext}

Recently, we started the study of a new class of theories for Majorana neutrinos~\cite{Debnath:2023akj}. In this study, we investigated the case where the fermionic dark matter is a Dirac fermion and using the cosmological bounds on the dark matter relic density we found an upper bound on the symmetry breaking scale in the multi-TeV energy scale. In Ref.~\cite{Debnath:2023akj} we provided a detailed analysis of these theories, we discussed the dark matter annihilation channels and the possibility to look for lepton number violating signatures from Higgs decays at the Large Hadron Collider. 

In this article, we study a simple theory where both, the Standard Model neutrinos and the new fermionic dark matter candidate, are Majorana fermions. We discuss in detail the theoretical framework, and the relation between the seesaw scale and the new symmetry breaking scale. We investigate all dark matter annihilation channels and point out the channels that are velocity suppressed. Since the dark matter is a Majorana fermion, it has a vector-axial coupling to the new gauge boson associated to the spontaneous breaking of the total lepton number, and several annihilation channels are velocity suppressed. This is a blessing because the relic density constraints imply that the upper bound on the mass of the new gauge boson has to be much smaller than in the case studied in Ref.~\cite{Debnath:2023akj}, where the dark matter is a Dirac fermion. In the most generic region of the parameter space (outside the $Z_l$-resonance) one predicts that the upper bound on the symmetry breaking scale is below $30$ TeV. Therefore, one can hope to test this theory at colliders.

This theory predicts a light pseudo-Nambu-Goldstone boson associated to the origin of neutrino masses, the Majoron~\cite{Chikashige:1980ui}. Since this new field is typically very light, one can achieve the correct dark matter relic density in regions of the parameter space where the new gauge boson is not very heavy. We study in detail the properties of the Majoron and point out that apart from the couplings to the Standard Model neutrinos, the couplings to photons is generated at one-loop level where inside the loop we have the new electrically charged fermionic fields needed for anomaly cancellation. We study the Majoron decays and investigate the impact of the cosmological bounds from Big Bang Nucleosynthesis, showing that one can rule out a large fraction of the parameter space for the Majoron mass and its lifetime. 

This article is organized as follows: In Sec.~\ref{Sec1} we discuss the theory for Majorana neutrinos based on local lepton number, we discuss the anomaly cancellation, the main features of the Higgs sector, and the connection between the canonical seesaw scale and the symmetry breaking scale. In Sec.~\ref{Sec2} we discuss the main features of the Majorana dark matter candidate, pointing out the velocity suppressed channels, and we show the allowed parameter space where each independent annihilation channel is contributing to the relic density. Including all annihilation channels, we find the upper bound on the symmetry breaking scale. We also discuss the direct detection bounds. In Sec.~\ref{Sec3} we discuss the main properties of the pseudo-Nambu-Goldstone boson and the bounds from Big Bang Nucleosynthesis. In Sec.~\ref{Sec4} we summarize our main results.
\section{LOCAL LEPTON NUMBER}
\label{Sec1}
In order to understand the origin of neutrino masses one can consider a simple theory for the spontaneous breaking of the total lepton number. This theory is based on the gauge symmetry~\cite{FileviezPerez:2011pt,Duerr:2013dza,FileviezPerez:2014lnj,Debnath:2023akj} 
$$SU(3)_C \otimes SU(2)_L \otimes U(1)_Y \otimes U(1)_{\ell},$$ 
where $U(1)_{\ell}$ is the gauge group for total lepton number. 
Here the total lepton number is defined in the standard way:
$\ell=\ell_e+\ell_\mu+\ell_\tau$. 
In this context, using the Standard Model leptonic fields, $\ell_L \sim (1,2,-1/2,1)$ and $e_R \sim (1,1,-1,1)$, one can estimate the different anomalies:
\begin{eqnarray*}
{\mathcal{A}}(SU(2)_L^2 U(1)_\ell) &=&3/2, \\
{\mathcal{A}}(U(1)_Y^2 U(1)_\ell) &=& -3/2, \\
{\mathcal{A}}(U(1)_Y U(1)_\ell^2) &=& 0,\\
{\mathcal{A}}(U(1)_\ell^3)=3,  &\text{and}& 
{\mathcal{A}} (U(1)_\ell)=3.
\end{eqnarray*}
The last two anomalies, ${\mathcal{A}}(U(1)_\ell^3)$ and ${\mathcal{A}} (U(1)_\ell)$, can be cancelled if one adds three copies of right-handed neutrinos, $\nu_R \sim (1,1,0,1)$, needed to implement the canonical seesaw mechanism. 

There are mainly two simple ways to cancel all anomalies listed above without spoiling the anomaly cancellation in the Standard Model:
\begin{itemize}
\item Adding vector-like leptons~\cite{FileviezPerez:2011pt,Duerr:2013dza}.
\item Adding four extra fermionic representations~\cite{FileviezPerez:2014lnj}.
\end{itemize}
In both cases one always predict a dark matter candidate from anomaly cancellation. In the first case, the dark matter candidate can be a Majorana or Dirac field, while in the second case, it has to be Majorana. 
In this article, we will investigate the case when the fermionic dark matter candidate is a Majorana fermion. For some studies in this context see Refs.~\cite{FileviezPerez:2011pt,Schwaller:2013hqa,FileviezPerez:2019cyn,Carena:2022qpf,Madge:2018gfl,FileviezPerez:2015mlm,Aranda:2014zta}.

The relevant Lagrangian for our discussion can be written as 
\begin{eqnarray}
\mathcal{L} &\supset & \hspace{0.2 cm} i \bar{\ell}_L \slashed D \ell_L + i \bar{e}_R \slashed D e_R+i \bar{\nu}_R \slashed D \nu_R + i \bar{\chi}_L \slashed D \chi_L \nonumber \\
&-& ( Y_e \ \bar{\ell}_L H e_R +Y_\nu \ \bar{\ell}_L i \sigma_2 H^* \nu_R \nonumber\\
& + & \lambda_R \ \nu^T_R C \nu_R \phi +\lambda_{\chi} \ \chi^T_L C \chi_L S^* + \text{h.c.}) \nonumber \\
&-& V(H,S,\phi) .
\label{lagrangian}
\end{eqnarray}
Here the $H$ is the Standard Model Higgs doublet which transforms as $H \sim (\mathbf{1},\mathbf{2},1/2,0)$, and the new Higgses, $S$ and $\phi$, transform as $S \sim (\mathbf{1},\mathbf{1},0,3)$ and $\phi \sim (\mathbf{1},\mathbf{1},0,-2)$. The field $\chi_L$ transforms as $\chi_L \sim (1,1,0,3/2)$.
In the above equation the covariant 
derivatives are defined as
\begin{eqnarray*}
  \slashed D \ell_L  &=& \slashed \partial \ell_L + i g_2 \slashed W \ell_L - i \frac{1}{2} g_1 \slashed B \ell_L + i g_\ell \slashed Z_\ell \ell_L, \\
 \slashed D e_R &=& \slashed \partial e_R  - i g_1 \slashed B e_R  + i g_\ell \slashed Z_\ell e_R,  \\
 \slashed D \nu_R &=& \slashed \partial \nu_R + i g_\ell \slashed Z_\ell \nu_R, \ \text{and} \\
 \slashed D \chi_L &=& \slashed \partial \chi_L + i n_\chi g_\ell \slashed Z_\ell \chi_L.
\end{eqnarray*}
Here $n_\chi=3/2$ as predicted by anomaly cancellation.
Once $U(1)_{\ell}$ is spontaneously broken, the Majorana dark matter candidate, $\chi=\chi_L + (\chi_L)^C$, stability is protected by a $\mathbb{Z}_2$ symmetry: $$\chi_L \to - \chi_L \ (\textrm{or} \ \chi \to - \chi).$$
Therefore, the stability of our dark matter candidate is a natural consequence from the spontaneous breaking of local lepton number.  

\begin{itemize}
%
\item Higgs Sector:
 The  most general scalar potential in this theory is given by 
 \begin{eqnarray}
 V(H,S,\phi)&=&-m_H^2 H ^{\dagger}H+\lambda(H^{\dagger}H)^2-m_s^2 S ^{\dagger}S \nonumber \\
 &+& \lambda_s (S^{\dagger}S)^2-m_{\phi}^2 \phi ^{\dagger}\phi 
 + \lambda_{\phi}(\phi^{\dagger}\phi)^2 \nonumber \\
 &+& \lambda_1(H^{\dagger}H)S^{\dagger}S + \lambda_2(H^{\dagger}H)\phi^{\dagger}\phi \nonumber \\
 &+& \lambda_3(S^{\dagger}S)\phi^{\dagger}\phi.
 \end{eqnarray}
The scalar fields in this theory can be written as 
\begin{eqnarray}
H&=&\begin{pmatrix}
h^+\\
\frac{1}{\sqrt{2}}(v_{0} + h_0) e^{i \sigma_0/v_0} 
\end{pmatrix}, \\
S &=& \frac{1}{\sqrt{2}}\left(v_{s} + h_s \right) e^{i \sigma_s/v_s}, 
\end{eqnarray}
and
\begin{eqnarray}
\phi &=& \frac{1}{\sqrt{2}}\left( v_{\phi} + h_{\phi} \right) e^{i \sigma_{\phi}/v_{\phi}} .
\end{eqnarray}
Notice that this scalar potential has the global symmetry: 
$O(4)_H \otimes U(1)_\phi \otimes U(1)_S $. 

In our notation the physical CP-even Higgses, $(h,H_1,H_2)$, are defined as
\begin{eqnarray}
\begin{pmatrix}
h_0\\
h_s\\
h_{\phi} 
\end{pmatrix} =
U
\begin{pmatrix}
h\\
H_1\\
H_2
\end{pmatrix}.
\label{Umixing}
\end{eqnarray}
There are three CP-odd Higgses and two of them are Goldstone's bosons eaten by the neutral gauge bosons. The gauge symmetry of this theory allows a five dimensional term in the potential:
\begin{eqnarray}
 V(H,S,\phi) \supset  \lambda_{M} \frac{S^2 \phi^3}{\Lambda}+ {\rm h.c.}. 
 \end{eqnarray}
 \\
 This term breaks the $U(1)_\phi \otimes U(1)_S$ symmetry of the potential and one gets a pseudo-Nambu-Goldstone boson, the Majoron $J$.
 The CP-odd Higgs eigenstates are defined by
\begin{eqnarray}
\begin{pmatrix}
\sigma_s \\
\sigma_{\phi} 
\end{pmatrix} =
\begin{pmatrix}
\cos \beta & \sin \beta \\
- \sin \beta & \cos \beta \\
\end{pmatrix}
\begin{pmatrix}
G_{\ell} \\
J
\end{pmatrix},
\end{eqnarray}
where 
\begin{equation}
\tan 2 \beta= \frac{12 v_s v_\phi}{4 v_\phi^2 - 9 v_s^2}.
\end{equation}
 For a detailed discussion of the Higgs sector see Ref.~\cite{Debnath:2023akj}.
 %
\item Canonical Seesaw:
One can generate Majorana neutrino masses using the interactions in Eq.(\ref{lagrangian}).
Thus the neutrino masses are generated through the type I seesaw mechanism and the Standard Model neutrino mass matrix is given by
\begin{equation}
    M_\nu = \frac{v_0^2}{2} Y_\nu M_N^{-1} Y_\nu^T,
\end{equation}
where 
\begin{equation}
M_N=\sqrt{2} \lambda_R v_\phi= \frac{\lambda_R}{\sqrt{2}} \frac{M_{Z_\ell}}{g_\ell} \cos \beta.
\label{MNZL}
\end{equation}
In the above equation, $v_0=246$ GeV, is the vacuum expectation value of the Standard Model Higgs and we used $v_\phi=v \cos \beta / 2$ and $v_s=v \sin \beta / 3$. One can write $M_{Z_\ell}=g_\ell \ v$. From Eq.~(\ref{MNZL}) one can see that the upper bound on the seesaw scale is mainly determined by the ratio $M_{Z_\ell}/ g_\ell$ and the perturbative bound on the Yukawa coupling $\lambda_R$.  
\end{itemize}
%
\section{MAJORANA DARK MATTER}
\label{Sec2}
\begin{figure}[h]
\centering
\includegraphics[width=0.45\textwidth]{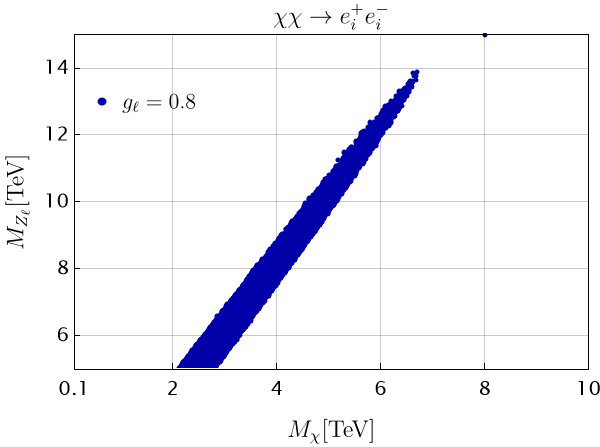}
\caption{Annihilation into charged leptons.}
\label{DMDMee}
\end{figure}

We pointed out that the dark matter candidate is a Majorana candidate: $\chi=\chi_L + (\chi_L)^C$.
\begin{widetext}
\begin{eqnarray*}
\begin{gathered}
\begin{tikzpicture}[line width=1.5 pt,node distance=1 cm and 1.5 cm]
\coordinate[label = left: $\chi$] (i1);
\coordinate[below right = 1cm of i1](v1);
\coordinate[below left = 1cm of v1, label= left:$\chi$](i2);
\coordinate[above right = 1cm of v1, label=right: $e_i$] (f1);
\coordinate[below right =  1cm of v1,label=right: $ e_i$] (f2);
\draw[fermionnoarrow] (i1) -- (v1);
\draw[fermionnoarrow] (i2) -- (v1);
\draw[fermion] (v1) -- (f1);
\draw[fermion] (f2) -- (v1);
\draw[fill=gray] (v1) circle (.3cm);
\end{tikzpicture}
\end{gathered} 
&=&
\begin{gathered}
\begin{tikzpicture}[line width=1.5 pt,node distance=1 cm and 1.5 cm]
\coordinate[label =left: $\chi$] (i1);
\coordinate[below right= 1cm of i1](v1);
\coordinate[ right= 0.5cm of v1,label= above:$Z_\ell$](vaux);
\coordinate[below left= 1cm of v1, label= left: $\chi$](i2);
\coordinate[right = 1 cm of v1](v2);
\coordinate[above right = 1 cm of v2, label=right: $e_i$] (f1);
\coordinate[below right =  1 cm of v2,label=right: $e_i$] (f2);
\draw[fermionnoarrow] (i1) -- (v1);
\draw[fermionnoarrow] (i2) -- (v1);
\draw[vector] (v1) -- (v2);
\draw[fermion] (v2) -- (f1);
\draw[fermion] (f2) -- (v2);
\draw[fill=cyan] (v1) circle (.1cm);
\draw[fill=cyan] (v2) circle (.1cm);
\end{tikzpicture}
\end{gathered} 
+
\begin{gathered}
\begin{tikzpicture}[line width=1.5 pt,node distance=1 cm and 1.5 cm]
\coordinate[label =left: $\chi$] (i1);
\coordinate[below right= 1cm of i1](v1);
\coordinate[ right= 0.5cm of v1,label= above:{$h$, $H_i$}](vaux);
\coordinate[below left= 1cm of v1, label= left: $\chi$](i2);
\coordinate[right = 1 cm of v1](v2);
\coordinate[above right = 1 cm of v2, label=right: $e_i$] (f1);
\coordinate[below right =  1 cm of v2,label=right: $e_i$] (f2);
\draw[fermionnoarrow] (i1) -- (v1);
\draw[fermionnoarrow] (i2) -- (v1);
\draw[scalarnoarrow] (v1) -- (v2);
\draw[fermion] (v2) -- (f1);
\draw[fermion] (f2) -- (v2);
\draw[fill=black] (v1) circle (.1cm);
\draw[fill=black] (v2) circle (.1cm);
\end{tikzpicture}
\end{gathered} \\
\begin{gathered}
\begin{tikzpicture}[line width=1.5 pt,node distance=1 cm and 1.5 cm]
\coordinate[label = left: $\chi$] (i1);
\coordinate[below right = 1cm of i1](v1);
\coordinate[below left = 1cm of v1, label= left:$\chi$](i2);
\coordinate[above right = 1cm of v1, label=right: $\nu$] (f1);
\coordinate[below right =  1cm of v1,label=right: $\nu$] (f2);
\draw[fermionnoarrow] (i1) -- (v1);
\draw[fermionnoarrow] (i2) -- (v1);
\draw[fermionnoarrow] (v1) -- (f1);
\draw[fermionnoarrow] (f2) -- (v1);
\draw[fill=gray] (v1) circle (.3cm);
\end{tikzpicture}
\end{gathered} 
&=&
\begin{gathered}
\begin{tikzpicture}[line width=1.5 pt,node distance=1 cm and 1.5 cm]
\coordinate[label =left: $\chi$] (i1);
\coordinate[below right= 1cm of i1](v1);
\coordinate[ right= 0.5cm of v1,label= above:$Z_\ell$](vaux);
\coordinate[below left= 1cm of v1, label= left: $\chi$](i2);
\coordinate[right = 1 cm of v1](v2);
\coordinate[above right = 1 cm of v2, label=right: $\nu$] (f1);
\coordinate[below right =  1 cm of v2,label=right: $\nu$] (f2);
\draw[fermionnoarrow] (i1) -- (v1);
\draw[fermionnoarrow] (i2) -- (v1);
\draw[vector] (v1) -- (v2);
\draw[fermionnoarrow] (v2) -- (f1);
\draw[fermionnoarrow] (f2) -- (v2);
\draw[fill=cyan] (v1) circle (.1cm);
\draw[fill=cyan] (v2) circle (.1cm);
\end{tikzpicture}
\end{gathered} 
+
\begin{gathered}
\begin{tikzpicture}[line width=1.5 pt,node distance=1 cm and 1.5 cm]
\coordinate[label =left: $\chi$] (i1);
\coordinate[below right= 1cm of i1](v1);
\coordinate[ right= 0.5cm of v1,label= above:{$J$}](vaux);
\coordinate[below left= 1cm of v1, label= left: $\chi$](i2);
\coordinate[right = 1 cm of v1](v2);
\coordinate[above right = 1 cm of v2, label=right: $\nu$] (f1);
\coordinate[below right =  1 cm of v2,label=right: $\nu$] (f2);
\draw[fermionnoarrow] (i1) -- (v1);
\draw[fermionnoarrow] (i2) -- (v1);
\draw[scalarnoarrow] (v1) -- (v2);
\draw[fermionnoarrow] (v2) -- (f1);
\draw[fermionnoarrow] (f2) -- (v2);
\draw[fill=black] (v1) circle (.1cm);
\draw[fill=black] (v2) circle (.1cm);
\end{tikzpicture}
\end{gathered} \\
\begin{gathered}
\begin{tikzpicture}[line width=1.5 pt,node distance=1 cm and 1.5 cm]
\coordinate[label = left: $\chi$] (i1);
\coordinate[below right = 1cm of i1](v1);
\coordinate[below left = 1cm of v1, label= left:$\chi$](i2);
\coordinate[above right = 1cm of v1, label=right: $N$] (f1);
\coordinate[below right =  1cm of v1,label=right: $N$] (f2);
\draw[fermionnoarrow] (i1) -- (v1);
\draw[fermionnoarrow] (i2) -- (v1);
\draw[fermionnoarrow] (v1) -- (f1);
\draw[fermionnoarrow] (f2) -- (v1);
\draw[fill=gray] (v1) circle (.3cm);
\end{tikzpicture}
\end{gathered} 
&=&
\begin{gathered}
\begin{tikzpicture}[line width=1.5 pt,node distance=1 cm and 1.5 cm]
\coordinate[label =left: $\chi$] (i1);
\coordinate[below right= 1cm of i1](v1);
\coordinate[ right= 0.5cm of v1,label= above:$J$](vaux);
\coordinate[below left= 1cm of v1, label= left: $\chi$](i2);
\coordinate[right = 1 cm of v1](v2);
\coordinate[above right = 1 cm of v2, label=right: $N$] (f1);
\coordinate[below right =  1 cm of v2,label=right: $N$] (f2);
\draw[fermionnoarrow] (i1) -- (v1);
\draw[fermionnoarrow] (i2) -- (v1);
\draw[scalarnoarrow] (v1) -- (v2);
\draw[fermionnoarrow] (v2) -- (f1);
\draw[fermionnoarrow] (f2) -- (v2);
\draw[fill=black] (v1) circle (.1cm);
\draw[fill=black] (v2) circle (.1cm);
\end{tikzpicture}
\end{gathered} 
+
\begin{gathered}
\begin{tikzpicture}[line width=1.5 pt,node distance=1 cm and 1.5 cm]
\coordinate[label =left: $\chi$] (i1);
\coordinate[below right= 1cm of i1](v1);
\coordinate[ right= 0.5cm of v1,label= above:{$Z_\ell$}](vaux);
\coordinate[below left= 1cm of v1, label= left: $\chi$](i2);
\coordinate[right = 1 cm of v1](v2);
\coordinate[above right = 1 cm of v2, label=right: $N$] (f1);
\coordinate[below right =  1 cm of v2,label=right: $N$] (f2);
\draw[fermionnoarrow] (i1) -- (v1);
\draw[fermionnoarrow] (i2) -- (v1);
\draw[vector] (v1) -- (v2);
\draw[fermionnoarrow] (v2) -- (f1);
\draw[fermionnoarrow] (f2) -- (v2);
\draw[fill=cyan] (v1) circle (.1cm);
\draw[fill=cyan] (v2) circle (.1cm);
\end{tikzpicture}
\end{gathered}
+
\begin{gathered}
\begin{tikzpicture}[line width=1.5 pt,node distance=1 cm and 1.5 cm]
\coordinate[label =left: $\chi$] (i1);
\coordinate[below right= 1cm of i1](v1);
\coordinate[ right= 0.5cm of v1,label= above:$H_i$](vaux);
\coordinate[below left= 1cm of v1, label= left: $\chi$](i2);
\coordinate[right = 1 cm of v1](v2);
\coordinate[above right = 1 cm of v2, label=right: $N$] (f1);
\coordinate[below right =  1 cm of v2,label=right: $N$] (f2);
\draw[fermionnoarrow] (i1) -- (v1);
\draw[fermionnoarrow] (i2) -- (v1);
\draw[scalarnoarrow] (v1) -- (v2);
\draw[fermionnoarrow] (v2) -- (f1);
\draw[fermionnoarrow] (f2) -- (v2);
\draw[fill=black] (v1) circle (.1cm);
\draw[fill=black] (v2) circle (.1cm);
\end{tikzpicture}
\end{gathered}\\
\begin{gathered}
\begin{tikzpicture}[line width=1.5 pt,node distance=1 cm and 1.5 cm]
\coordinate[label = left: $\chi$] (i1);
\coordinate[below right = 1cm of i1](v1);
\coordinate[below left = 1cm of v1, label= left:$\chi$](i2);
\coordinate[above right = 1cm of v1, label=right: $Z_\ell$] (f1);
\coordinate[below right =  1cm of v1,label=right: $Z_\ell$] (f2);
\draw[fermionnoarrow] (i1) -- (v1);
\draw[fermionnoarrow] (i2) -- (v1);
\draw[vector] (v1) -- (f1);
\draw[vector] (f2) -- (v1);
\draw[fill=gray] (v1) circle (.3cm);
\end{tikzpicture}
\end{gathered} 
&=&
\begin{gathered}
\begin{tikzpicture}[line width=1.5 pt,node distance=1 cm and 1.5 cm]
\coordinate[label =left: $\chi$] (i1);
\coordinate[right= 1cm of i1](v1);
\coordinate[below= 0.5cm of v1](vaux);
\coordinate[right = 1cm of v1, label= right:$Z_\ell$](f1);
\coordinate[below = 1 cm of v1](v2);
\coordinate[left = 1 cm of v2, label=left: $\chi$] (i2);
\coordinate[right =  1 cm of v2,label=right: $Z_\ell$] (f2);
\draw[fermionnoarrow] (i1) -- (v1);
\draw[fermionnoarrow] (v1) -- (v2);
\draw[fermionnoarrow] (i2) -- (v2);
\draw[vector] (f2) -- (v2);
\draw[vector] (f1) -- (v1);
\draw[fill=cyan] (v1) circle (.1cm);
\draw[fill=cyan] (v2) circle (.1cm);
\end{tikzpicture}
\end{gathered}
\quad   \! +
\begin{gathered}
\begin{tikzpicture}[line width=1.5 pt,node distance=1 cm and 1.5 cm]
\coordinate[label =left: $\chi$] (i1);
\coordinate[right= 1cm of i1](v1);
\coordinate[below= 0.5cm of v1,label](vaux);
\coordinate[right = 1cm of v1, label= right:$Z_\ell$](f1);
\coordinate[below = 1 cm of v1](v2);
\coordinate[left = 1 cm of v2, label=left: $\chi$] (i2);
\coordinate[right =  1 cm of v2,label=right: $Z_\ell$] (f2);
\draw[fermionnoarrow] (i1) -- (v1);
\draw[fermionnoarrow] (v1) -- (v2);
\draw[fermionnoarrow] (i2) -- (v2);
\draw[vector] (f2) -- (v1);
\draw[vector] (f1) -- (v2);
\draw[fill=cyan] (v1) circle (.1cm);
\draw[fill=cyan] (v2) circle (.1cm);
\end{tikzpicture}
\end{gathered}
+
\begin{gathered}
\begin{tikzpicture}[line width=1.5 pt,node distance=1 cm and 1.5 cm]
\coordinate[label =left: $\chi$] (i1);
\coordinate[below right= 1cm of i1](v1);
\coordinate[ right= 0.5cm of v1,label= above:{$H_i$}](vaux);
\coordinate[below left= 1cm of v1, label= left: $\chi$](i2);
\coordinate[right = 1 cm of v1](v2);
\coordinate[above right = 1 cm of v2, label=right: $Z_\ell$] (f1);
\coordinate[below right =  1 cm of v2,label=right: $Z_\ell$] (f2);
\draw[fermionnoarrow] (i1) -- (v1);
\draw[fermionnoarrow] (i2) -- (v1);
\draw[scalarnoarrow] (v1) -- (v2);
\draw[vector] (v2) -- (f1);
\draw[vector] (f2) -- (v2);
\draw[fill=black] (v1) circle (.1cm);
\draw[fill=cyan] (v2) circle (.1cm);
\end{tikzpicture}
\end{gathered}
\\
\begin{gathered}
\begin{tikzpicture}[line width=1.5 pt,node distance=1 cm and 1.5 cm]
\coordinate[label = left: $\chi$] (i1);
\coordinate[below right = 1cm of i1](v1);
\coordinate[below left = 1cm of v1, label= left:$\chi$](i2);
\coordinate[above right = 1cm of v1, label=right: $Z_\ell$] (f1);
\coordinate[below right =  1cm of v1,label=right: $H_i$] (f2);
\draw[fermionnoarrow] (i1) -- (v1);
\draw[fermionnoarrow] (i2) -- (v1);
\draw[vector] (v1) -- (f1);
\draw[scalarnoarrow] (f2) -- (v1);
\draw[fill=gray] (v1) circle (.3cm);
\end{tikzpicture}
\end{gathered} 
&=&
\begin{gathered}
\begin{tikzpicture}[line width=1.5 pt,node distance=1 cm and 1.5 cm]
\coordinate[label =left: $\chi$] (i1);
\coordinate[right= 1cm of i1](v1);
\coordinate[below= 0.5cm of v1](vaux);
\coordinate[right = 1cm of v1, label= right:$Z_\ell$](f1);
\coordinate[below = 1 cm of v1](v2);
\coordinate[left = 1 cm of v2, label=left: $\chi$] (i2);
\coordinate[right =  1 cm of v2,label=right: $H_i$] (f2);
\draw[fermionnoarrow] (i1) -- (v1);
\draw[fermionnoarrow] (v1) -- (v2);
\draw[fermionnoarrow] (i2) -- (v2);
\draw[scalarnoarrow] (f2) -- (v2);
\draw[vector] (f1) -- (v1);
\draw[fill=cyan] (v1) circle (.1cm);
\draw[fill=black] (v2) circle (.1cm);
\end{tikzpicture}
\end{gathered}
\quad   \! +
\begin{gathered}
\begin{tikzpicture}[line width=1.5 pt,node distance=1 cm and 1.5 cm]
\coordinate[label =left: $\chi$] (i1);
\coordinate[right= 1cm of i1](v1);
\coordinate[below= 0.5cm of v1,label](vaux);
\coordinate[right = 1cm of v1, label= right:$Z_\ell$](f1);
\coordinate[below = 1 cm of v1](v2);
\coordinate[left = 1 cm of v2, label=left: $\chi$] (i2);
\coordinate[right =  1 cm of v2,label=right: $H_i$] (f2);
\draw[fermionnoarrow] (i1) -- (v1);
\draw[fermionnoarrow] (v1) -- (v2);
\draw[fermionnoarrow] (i2) -- (v2);
\draw[scalarnoarrow] (f2) -- (v1);
\draw[vector] (f1) -- (v2);
\draw[fill=black] (v1) circle (.1cm);
\draw[fill=cyan] (v2) circle (.1cm);
\end{tikzpicture}
\end{gathered}
+
\begin{gathered}
\begin{tikzpicture}[line width=1.5 pt,node distance=1 cm and 1.5 cm]
\coordinate[label =left: $\chi$] (i1);
\coordinate[below right= 1cm of i1](v1);
\coordinate[ right= 0.5cm of v1,label= above:{$Z_\ell$}](vaux);
\coordinate[below left= 1cm of v1, label= left: $\chi$](i2);
\coordinate[right = 1 cm of v1](v2);
\coordinate[above right = 1 cm of v2, label=right: $Z_\ell$] (f1);
\coordinate[below right =  1 cm of v2,label=right: $H_i$] (f2);
\draw[fermionnoarrow] (i1) -- (v1);
\draw[fermionnoarrow] (i2) -- (v1);
\draw[vector] (v1) -- (v2);
\draw[vector] (v2) -- (f1);
\draw[scalarnoarrow] (f2) -- (v2);
\draw[fill=cyan] (v1) circle (.1cm);
\draw[fill=cyan] (v2) circle (.1cm);
\end{tikzpicture}
\end{gathered}
\\
\begin{gathered}
\begin{tikzpicture}[line width=1.5 pt,node distance=1 cm and 1.5 cm]
\coordinate[label = left: $\chi$] (i1);
\coordinate[below right = 1cm of i1](v1);
\coordinate[below left = 1cm of v1, label= left:$\chi$](i2);
\coordinate[above right = 1cm of v1, label=right: $J$] (f1);
\coordinate[below right =  1cm of v1,label=right: $J$] (f2);
\draw[fermionnoarrow] (i1) -- (v1);
\draw[fermionnoarrow] (i2) -- (v1);
\draw[scalarnoarrow] (v1) -- (f1);
\draw[scalarnoarrow] (f2) -- (v1);
\draw[fill=gray] (v1) circle (.3cm);
\end{tikzpicture}
\end{gathered} 
&=& 
\begin{gathered}
\begin{tikzpicture}[line width=1.5 pt,node distance=1 cm and 1.5 cm]
\coordinate[label =left: $\chi$] (i1);
\coordinate[right= 1cm of i1](v1);
\coordinate[below= 0.5cm of v1](vaux);
\coordinate[right = 1cm of v1, label= right:$J$](f1);
\coordinate[below = 1 cm of v1](v2);
\coordinate[left = 1 cm of v2, label=left: $\chi$] (i2);
\coordinate[right =  1 cm of v2,label=right: $J$] (f2);
\draw[fermionnoarrow] (i1) -- (v1);
\draw[fermionnoarrow] (v1) -- (v2);
\draw[fermionnoarrow] (i2) -- (v2);
\draw[scalarnoarrow] (f2) -- (v2);
\draw[scalarnoarrow] (f1) -- (v1);
\draw[fill=black] (v1) circle (.1cm);
\draw[fill=black] (v2) circle (.1cm);
\end{tikzpicture}
\end{gathered}
\quad   \! +
\begin{gathered}
\begin{tikzpicture}[line width=1.5 pt,node distance=1 cm and 1.5 cm]
\coordinate[label =left: $\chi$] (i1);
\coordinate[right= 1cm of i1](v1);
\coordinate[below= 0.5cm of v1,label](vaux);
\coordinate[right = 1cm of v1, label= right:$J$](f1);
\coordinate[below = 1 cm of v1](v2);
\coordinate[left = 1 cm of v2, label=left: $\chi$] (i2);
\coordinate[right =  1 cm of v2,label=right: $J$] (f2);
\draw[fermionnoarrow] (i1) -- (v1);
\draw[fermionnoarrow] (v1) -- (v2);
\draw[fermionnoarrow] (i2) -- (v2);
\draw[scalarnoarrow] (f2) -- (v1);
\draw[scalarnoarrow] (f1) -- (v2);
\draw[fill=black] (v1) circle (.1cm);
\draw[fill=black] (v2) circle (.1cm);
\end{tikzpicture}
\end{gathered}
\\
\begin{gathered}
\begin{tikzpicture}[line width=1.5 pt,node distance=1 cm and 1.5 cm]
\coordinate[label = left: $\chi$] (i1);
\coordinate[below right = 1cm of i1](v1);
\coordinate[below left = 1cm of v1, label= left:$\chi$](i2);
\coordinate[above right = 1cm of v1, label=right: $H_i$] (f1);
\coordinate[below right =  1cm of v1,label=right: $H_j$] (f2);
\draw[fermionnoarrow] (i1) -- (v1);
\draw[fermionnoarrow] (i2) -- (v1);
\draw[scalarnoarrow] (v1) -- (f1);
\draw[scalarnoarrow] (f2) -- (v1);
\draw[fill=gray] (v1) circle (.3cm);
\end{tikzpicture}
\end{gathered} 
&=&
\begin{gathered}
\begin{tikzpicture}[line width=1.5 pt,node distance=1 cm and 1.5 cm]
\coordinate[label =left: $\chi$] (i1);
\coordinate[right= 1cm of i1](v1);
\coordinate[below= 0.5cm of v1](vaux);
\coordinate[right = 1cm of v1, label= right:$H_i$](f1);
\coordinate[below = 1 cm of v1](v2);
\coordinate[left = 1 cm of v2, label=left: $\chi$] (i2);
\coordinate[right =  1 cm of v2,label=right: $H_j$] (f2);
\draw[fermionnoarrow] (i1) -- (v1);
\draw[fermionnoarrow] (v1) -- (v2);
\draw[fermionnoarrow] (i2) -- (v2);
\draw[scalarnoarrow] (f2) -- (v2);
\draw[scalarnoarrow] (f1) -- (v1);
\draw[fill=black] (v1) circle (.1cm);
\draw[fill=black] (v2) circle (.1cm);
\end{tikzpicture}
\end{gathered}
\quad   \! +
\begin{gathered}
\begin{tikzpicture}[line width=1.5 pt,node distance=1 cm and 1.5 cm]
\coordinate[label =left: $\chi$] (i1);
\coordinate[right= 1cm of i1](v1);
\coordinate[below= 0.5cm of v1,label](vaux);
\coordinate[right = 1cm of v1, label= right:$H_i$](f1);
\coordinate[below = 1 cm of v1](v2);
\coordinate[left = 1 cm of v2, label=left: $\chi$] (i2);
\coordinate[right =  1 cm of v2,label=right: $H_j$] (f2);
\draw[fermionnoarrow] (i1) -- (v1);
\draw[fermionnoarrow] (v1) -- (v2);
\draw[fermionnoarrow] (i2) -- (v2);
\draw[scalarnoarrow] (f2) -- (v1);
\draw[scalarnoarrow] (f1) -- (v2);
\draw[fill=black] (v1) circle (.1cm);
\draw[fill=black] (v2) circle (.1cm);
\end{tikzpicture}
\end{gathered}
+
\begin{gathered}
\begin{tikzpicture}[line width=1.5 pt,node distance=1 cm and 1.5 cm]
\coordinate[label =left: $\chi$] (i1);
\coordinate[below right= 1cm of i1](v1);
\coordinate[ right= 0.5cm of v1,label= above:{$h$,$H_k$}](vaux);
\coordinate[below left= 1cm of v1, label= left: $\chi$](i2);
\coordinate[right = 1 cm of v1](v2);
\coordinate[above right = 1 cm of v2, label=right: $H_i$] (f1);
\coordinate[below right =  1 cm of v2,label=right: $H_j$] (f2);
\draw[fermionnoarrow] (i1) -- (v1);
\draw[fermionnoarrow] (i2) -- (v1);
\draw[scalarnoarrow] (v1) -- (v2);
\draw[scalarnoarrow] (v2) -- (f1);
\draw[scalarnoarrow] (f2) -- (v2);
\draw[fill=black] (v1) circle (.1cm);
\draw[fill=black] (v2) circle (.1cm);
\end{tikzpicture}
\end{gathered}\\
\begin{gathered}
\begin{tikzpicture}[line width=1.5 pt,node distance=1 cm and 1.5 cm]
\coordinate[label = left: $\chi$] (i1);
\coordinate[below right = 1cm of i1](v1);
\coordinate[below left = 1cm of v1, label= left:$\chi$](i2);
\coordinate[above right = 1cm of v1, label=right: $Z_\ell$] (f1);
\coordinate[below right =  1cm of v1,label=right: $J$] (f2);
\draw[fermionnoarrow] (i1) -- (v1);
\draw[fermionnoarrow] (i2) -- (v1);
\draw[vector] (v1) -- (f1);
\draw[scalarnoarrow] (f2) -- (v1);
\draw[fill=gray] (v1) circle (.3cm);
\end{tikzpicture}
\end{gathered} 
&=&
\begin{gathered}
\begin{tikzpicture}[line width=1.5 pt,node distance=1 cm and 1.5 cm]
\coordinate[label =left: $\chi$] (i1);
\coordinate[right= 1cm of i1](v1);
\coordinate[below= 0.5cm of v1](vaux);
\coordinate[right = 1cm of v1, label= right:$Z_\ell$](f1);
\coordinate[below = 1 cm of v1](v2);
\coordinate[left = 1 cm of v2, label=left: $\chi$] (i2);
\coordinate[right =  1 cm of v2,label=right: $J$] (f2);
\draw[fermionnoarrow] (i1) -- (v1);
\draw[fermionnoarrow] (v1) -- (v2);
\draw[fermionnoarrow] (i2) -- (v2);
\draw[scalarnoarrow] (f2) -- (v2);
\draw[vector] (f1) -- (v1);
\draw[fill=cyan] (v1) circle (.1cm);
\draw[fill=black] (v2) circle (.1cm);
\end{tikzpicture}
\end{gathered}
\quad   \! +
\begin{gathered}
\begin{tikzpicture}[line width=1.5 pt,node distance=1 cm and 1.5 cm]
\coordinate[label =left: $\chi$] (i1);
\coordinate[right= 1cm of i1](v1);
\coordinate[below= 0.5cm of v1,label](vaux);
\coordinate[right = 1cm of v1, label= right:$Z_\ell$](f1);
\coordinate[below = 1 cm of v1](v2);
\coordinate[left = 1 cm of v2, label=left: $\chi$] (i2);
\coordinate[right =  1 cm of v2,label=right: $J$] (f2);
\draw[fermionnoarrow] (i1) -- (v1);
\draw[fermionnoarrow] (v1) -- (v2);
\draw[fermionnoarrow] (i2) -- (v2);
\draw[scalarnoarrow] (f2) -- (v1);
\draw[vector] (f1) -- (v2);
\draw[fill=black] (v1) circle (.1cm);
\draw[fill=cyan] (v2) circle (.1cm);
\end{tikzpicture}
\end{gathered}
+
\begin{gathered}
\begin{tikzpicture}[line width=1.5 pt,node distance=1 cm and 1.5 cm]
\coordinate[label =left: $\chi$] (i1);
\coordinate[below right= 1cm of i1](v1);
\coordinate[ right= 0.5cm of v1,label= above:$H_i$](vaux);
\coordinate[below left= 1cm of v1, label= left: $\chi$](i2);
\coordinate[right = 1 cm of v1](v2);
\coordinate[above right = 1 cm of v2, label=right: $Z_\ell$] (f1);
\coordinate[below right =  1 cm of v2,label=right: $J$] (f2);
\draw[fermionnoarrow] (i1) -- (v1);
\draw[fermionnoarrow] (i2) -- (v1);
\draw[scalarnoarrow] (v1) -- (v2);
\draw[vector] (v2) -- (f1);
\draw[scalarnoarrow] (f2) -- (v2);
\draw[fill=black] (v1) circle (.1cm);
\draw[fill=cyan] (v2) circle (.1cm);
\end{tikzpicture}
\end{gathered}\\
\begin{gathered}
\begin{tikzpicture}[line width=1.5 pt,node distance=1 cm and 1.5 cm]
\coordinate[label = left: $\chi$] (i1);
\coordinate[below right = 1cm of i1](v1);
\coordinate[below left = 1cm of v1, label= left:$\chi$](i2);
\coordinate[above right = 1cm of v1, label=right: $H_i$] (f1);
\coordinate[below right =  1cm of v1,label=right: $J$] (f2);
\draw[fermionnoarrow] (i1) -- (v1);
\draw[fermionnoarrow] (i2) -- (v1);
\draw[scalarnoarrow] (v1) -- (f1);
\draw[scalarnoarrow] (f2) -- (v1);
\draw[fill=gray] (v1) circle (.3cm);
\end{tikzpicture}
\end{gathered} 
&=&
\begin{gathered}
\begin{tikzpicture}[line width=1.5 pt,node distance=1 cm and 1.5 cm]
\coordinate[label =left: $\chi$] (i1);
\coordinate[right= 1cm of i1](v1);
\coordinate[below= 0.5cm of v1](vaux);
\coordinate[right = 1cm of v1, label= right:$H_i$](f1);
\coordinate[below = 1 cm of v1](v2);
\coordinate[left = 1 cm of v2, label=left: $\chi$] (i2);
\coordinate[right =  1 cm of v2,label=right: $J$] (f2);
\draw[fermionnoarrow] (i1) -- (v1);
\draw[fermionnoarrow] (v1) -- (v2);
\draw[fermionnoarrow] (i2) -- (v2);
\draw[scalarnoarrow] (f2) -- (v2);
\draw[scalarnoarrow] (f1) -- (v1);
\draw[fill=black] (v1) circle (.1cm);
\draw[fill=black] (v2) circle (.1cm);
\end{tikzpicture}
\end{gathered}
\quad   \! +
\begin{gathered}
\begin{tikzpicture}[line width=1.5 pt,node distance=1 cm and 1.5 cm]
\coordinate[label =left: $\chi$] (i1);
\coordinate[right= 1cm of i1](v1);
\coordinate[below= 0.5cm of v1,label](vaux);
\coordinate[right = 1cm of v1, label= right:$H_i$](f1);
\coordinate[below = 1 cm of v1](v2);
\coordinate[left = 1 cm of v2, label=left: $\chi$] (i2);
\coordinate[right =  1 cm of v2,label=right: $J$] (f2);
\draw[fermionnoarrow] (i1) -- (v1);
\draw[fermionnoarrow] (v1) -- (v2);
\draw[fermionnoarrow] (i2) -- (v2);
\draw[scalarnoarrow] (f2) -- (v1);
\draw[scalarnoarrow] (f1) -- (v2);
\draw[fill=black] (v1) circle (.1cm);
\draw[fill=black] (v2) circle (.1cm);
\end{tikzpicture}
\end{gathered}
+
\begin{gathered}
\begin{tikzpicture}[line width=1.5 pt,node distance=1 cm and 1.5 cm]
\coordinate[label =left: $\chi$] (i1);
\coordinate[below right= 1cm of i1](v1);
\coordinate[ right= 0.5cm of v1,label= above:$Z_\ell$](vaux);
\coordinate[below left= 1cm of v1, label= left: $\chi$](i2);
\coordinate[right = 1 cm of v1](v2);
\coordinate[above right = 1 cm of v2, label=right: $H_i$] (f1);
\coordinate[below right =  1 cm of v2,label=right: $J$] (f2);
\draw[fermionnoarrow] (i1) -- (v1);
\draw[fermionnoarrow] (i2) -- (v1);
\draw[vector] (v1) -- (v2);
\draw[scalarnoarrow] (v2) -- (f1);
\draw[scalarnoarrow] (f2) -- (v2);
\draw[fill=cyan] (v1) circle (.1cm);
\draw[fill=cyan] (v2) circle (.1cm);
\end{tikzpicture}
\end{gathered}
\label{graphs}
\end{eqnarray*}
\end{widetext}
\pagestyle{plain}
\begin{table}
    \centering
    \begin{tabular}{||c|c||}
    \hline
     \text{Channel} & $v$-\text{Suppression} \\
     \hline\hline
      $\chi \chi \rightarrow \bar{e_i}e_i$  & yes \\
      \hline
      $\chi \chi \rightarrow \nu \nu$  & yes \\
      \hline
      $\chi \chi \rightarrow Z_\ell Z_\ell$   & no \\
      \hline
      $\chi \chi \rightarrow Z_\ell H_i$   & no \\
      \hline
      $\chi \chi \rightarrow JJ$   & yes \\ 
      \hline
      $\chi \chi \rightarrow H_i H_j$   & no \\
      \hline
      $\chi \chi \rightarrow NN$   & no \\
      \hline
      $\chi \chi \rightarrow Z_\ell J$  & no \\
      \hline
      $\chi \chi \rightarrow H_i J$ & no \\
      \hline
      $\chi \chi \rightarrow ZZ$  & yes \\
      \hline
      $\chi \chi \rightarrow \bar{q}q$  & yes \\
      \hline
      $\chi \chi \rightarrow W^+W^-$  & yes  \\
      \hline
    \end{tabular}
    \caption{Classification of dark matter annihilation channels according to their velocity dependence. Here we work in the limit where we neglect the SM neutrino, charged lepton and Majoron masses.}
    \label{clasiffication}
\end{table}
The main dark matter annihilation channels in this model are:
\begin{eqnarray*}
\chi \chi &\to& Z_\ell Z_\ell, \bar{e_i}e_i, \nu \nu, N N, J J, H_i H_j, \nonumber \\
& & Z_\ell H_i, Z_\ell J, H_i J, Z Z, W^+ W^-, q \bar{q}. \nonumber 
\end{eqnarray*}
The Feynman graphs for the most important channels are listed in the Figure above.
Notice the presence of the light pseudo-Nambu-Goldstone boson, the Majoron, which allow us to obtain the correct relic density in a large fraction of the parameter space. Since our dark matter candidate is a Majorana fermion, most of the annihilation channels through the $Z_\ell$ gauge boson are velocity suppressed. See Table.~\ref{clasiffication} for the classification of the dark matter annihilation channels according to their velocity dependence. 
The  relevant parameters in this model for the relic density calculation are 
$$g_{\ell}, M_\chi, M_{H_i},M_{Z_{\ell}}, M_N, M_J, \sin{\beta} \hspace{0.1 cm} \text{and} \hspace{0.1 cm} U_{ij}.$$
See appendix~\ref{AppendixA} for the Feynman rules.
Performing the standard freeze-out calculation for the dark matter relic density~\cite{Gondolo:1990dk}, in Fig.~\ref{DMDMee} we show the allowed parameter space by the cosmological bounds, $\Omega_\chi h^2 \leq 0.12$~\cite{Planck:2018vyg}, when our dark matter candidate annihilates only into the Standard Model charged leptons. Here we use $g_\ell=0.8$ which is basically the upper bound on the gauge coupling in this model. 
The maximal value of $g_\ell$ is determined by the perturbative bound on the $S^\dagger S Z_\ell Z_\ell$ coupling, which gives us that $g_\ell \leq 0.83$.
As one can appreciate, we obtain the correct relic density only when the process, $\chi \chi \to Z_\ell^* \to e_i^+ e_i^-$, occurs through the $Z_\ell$-resonance.
\begin{figure}[h]
\centering
\includegraphics[width=0.45\textwidth]{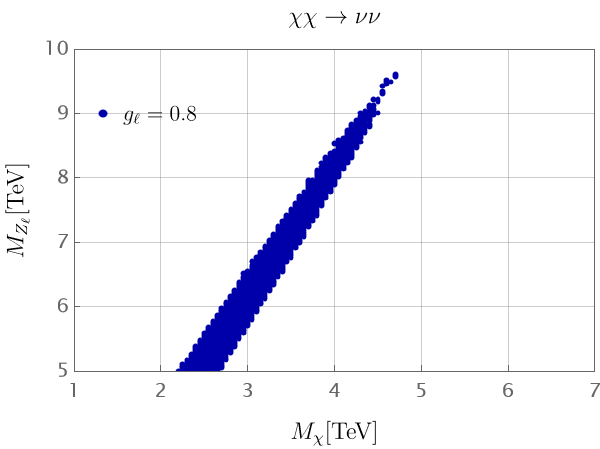}
\caption{Annihilation into the SM neutrinos.}
\label{DMDMnunu}
\end{figure}
In Fig.~\ref{DMDMnunu} we see a similar situation, but in this case we have only the annihilation into neutrinos, i.e $\chi \chi \to Z_\ell^* \to \nu \nu$. In both cases showed in Figs.~\ref{DMDMee} and \ref{DMDMnunu}, the annihilation cross sections are velocity suppressed.
\begin{figure}[h]
         \centering        \includegraphics[width=0.45\textwidth]{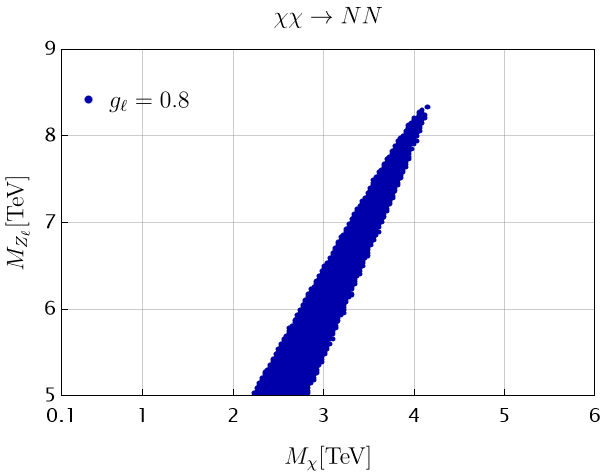}
         \caption{Annihilation into right-handed neutrinos.}
         \label{DMDMNN}
\end{figure}
   
In Fig.~\ref{DMDMNN} we show the allowed parameter space when the dark matter annihilates only into the right-handed neutrinos assuming that their masses are $1$ TeV. In this case one can achieve the correct relic density when $M_\chi \sim M_{Z_\ell}/2$. i.e. in the resonance region.   
 \begin{figure} [h]  
        \centering       \includegraphics[width=0.45\textwidth]{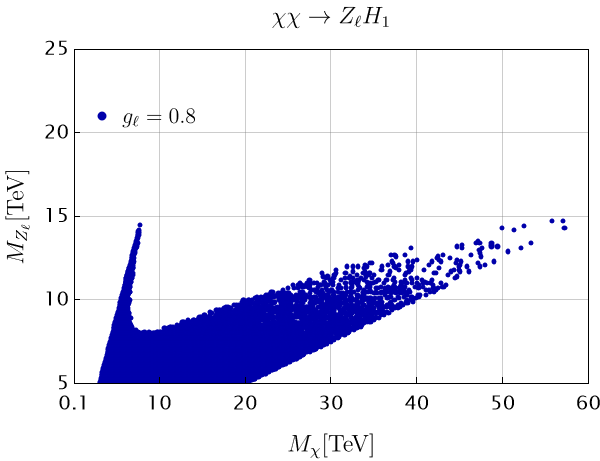}
         \caption{Annihilation into $Z_\ell H_1$.}
         \label{DMDMZH1}
\end{figure}
In Fig.~\ref{DMDMZH1} we show the viable parameter space to achieve the correct relic density when one has $\chi \chi \to Z_\ell H_1$. As one can see, there two main regions, the $Z_\ell$-resonance region and the other region where $M_\chi > M_{Z_\ell}/2$, which is more generic. For illustration, we assume that the $H_1$ mass is $1$ TeV. 
\begin{figure}[h]
         \centering
         \includegraphics[width=0.45\textwidth]{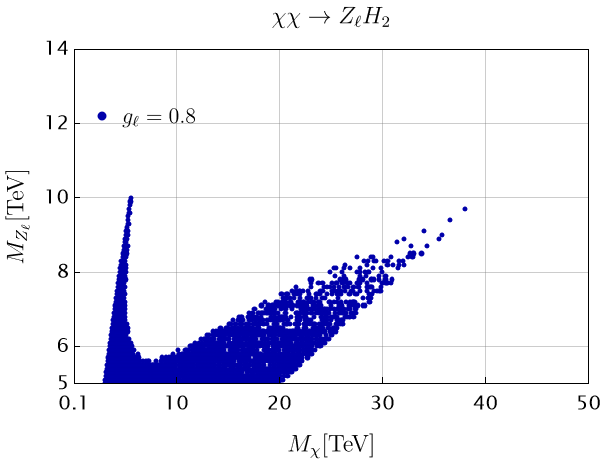}
         \caption{Annihilation into $Z_\ell H_2$.}
         \label{DMDMZH2}
\end{figure}
In Fig.~\ref{DMDMZH2} we show similar results as in Fig.~\ref{DMDMZH1}. In this case the dark matter annihilates into the $Z_\ell$ and the second new Higgs $H_2$ with mass $1$ TeV.
 \begin{figure}[h]  
\centering
\includegraphics[width=0.45\textwidth]{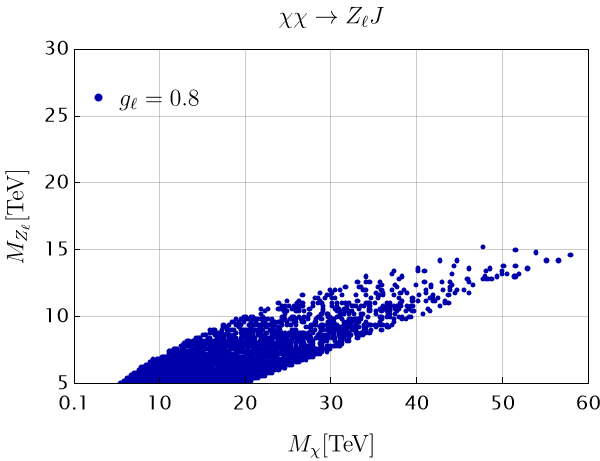}
\caption{Annihilation into $Z_\ell J$.}
\label{DMDMZJ}
\end{figure}
As in the previous case, the regions allowed by the relic density constraints have similar features, the only difference is basically the maximal allowed value for the gauge boson mass. In Fig.~\ref{DMDMZH2} the $Z_\ell$ mass is always below $10$ TeV, while in Fig.~\ref{DMDMZH1} one can reach $15$ TeV.

The existence of the Majoron opens up new possibilities for dark matter annihilation. In Fig.~\ref{DMDMZJ} we show the results for the allowed parameter space when one has only the annihilation, $\chi \chi \to Z_\ell J$. As one can appreciate, in this case, one can achieve the correct relic density without the need of a resonance and generically the $Z_\ell$ does not need to be very heavy.
\begin{figure}[h]
\centering
\includegraphics[width=0.45\textwidth]{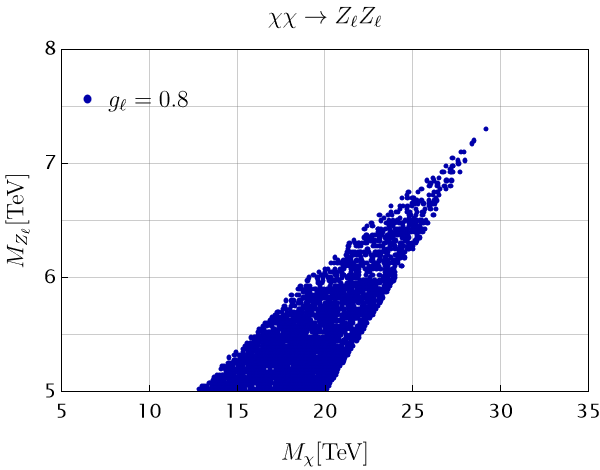}
\caption{Annihilation into $Z_\ell Z_\ell$.}
\label{DMDMZZ}
\end{figure}
\begin{figure}[h]  
\centering
\includegraphics[width=0.45\textwidth]{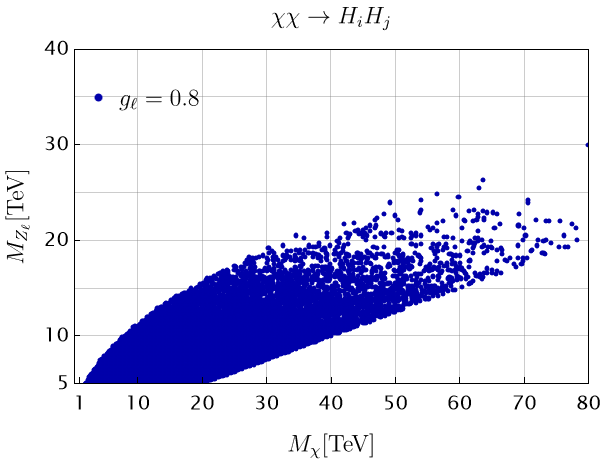}
\caption{Annihilation into $H_i H_j$.}
\label{DMDMHH}
\end{figure}
In Fig.~\ref{DMDMZZ} we show similar results when one has the annihilation into two new gauge bosons. In this case the annihilation cross section is also not velocity suppressed, one can easily achieve the correct relic density and the gauge boson mass is always below $8$ TeV.
In all these results we are imposing the collider bounds on the new gauge boson, i.e. $M_{Z_\ell}/g_\ell > (6-7)$ TeV~\cite{Alioli:2017nzr}.
The allowed parameter space when the dark matter annihilates only into the new two Higgses is shown in Fig.~\ref{DMDMHH}.
Since this annihilation channels are not velocity suppressed one can achieve the correct relic density even when the gauge boson is heavy, with mass below approximately $25$ TeV. 
\begin{figure}[h]
\centering
\includegraphics[width=0.45\textwidth]{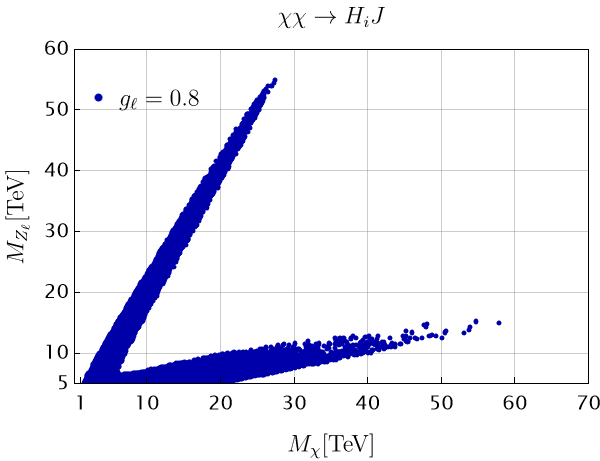}
\caption{Annihilation into $H_i J$.}
\label{DMDMHJ}
\end{figure}
\begin{figure}[h]
\centering
\includegraphics[width=0.45\textwidth]{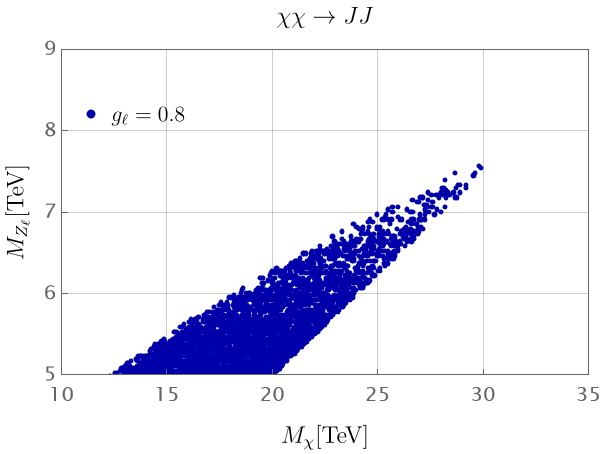}
\caption{Annihilation into $JJ$.}
\label{DMDMJJ}
\end{figure}
There are two viable regions when the dark matter annihilates into the new Higgses and the Majoron, see Fig.~\ref{DMDMHJ}.
Since this annihilation cross-section is not velocity suppressed the $Z_\ell$ mass can be large when one has the $Z_\ell$-resonance. The last channel we have is $\chi \chi \to JJ$ which is velocity suppressed and we show the results in Fig.~\ref{DMDMJJ}. In this case the allowed values for the $Z_\ell$-mass is always below $8$ TeV. The annihilation channels $\chi \chi \rightarrow W W, Z Z, q\bar{q}$ are velocity suppressed and their contribution to the relic density is negligible compared to other annihilation channels. Therefore, we do not include these channels in our discussion.

The previous discussion is very useful to identify the key properties of each dark matter annihilation channel, and then we can now show the allowed parameter space including all annihilation channels. In appendix~\ref{AppendixB} we also show the branching ratios for the different annihilation channels to understand the contribution of each channel in different regions of the parameter space.

\begin{figure}[t]  
\centering
\includegraphics[width=0.45\textwidth]{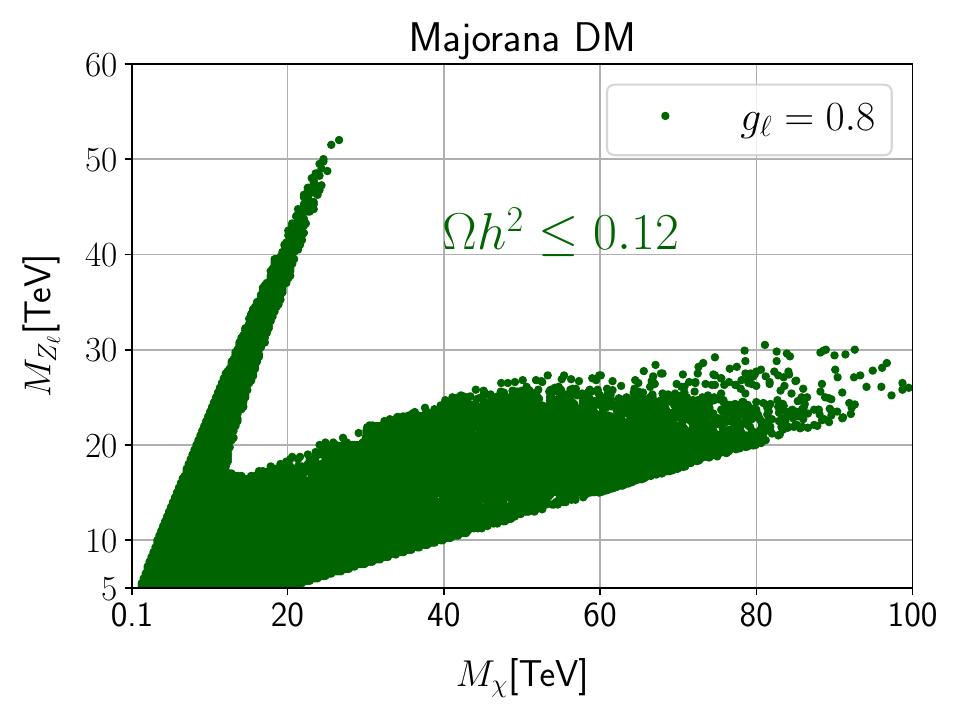}
\caption{Allowed parameter space by the relic density constraints when $g_\ell=0.8$. The area below the allowed region is excluded once we impose the perturbative bound on the Yukawa coupling $\lambda_\chi$.}
\label{all08}
\end{figure}
\begin{figure}[h]
\centering
\includegraphics[width=0.45\textwidth]{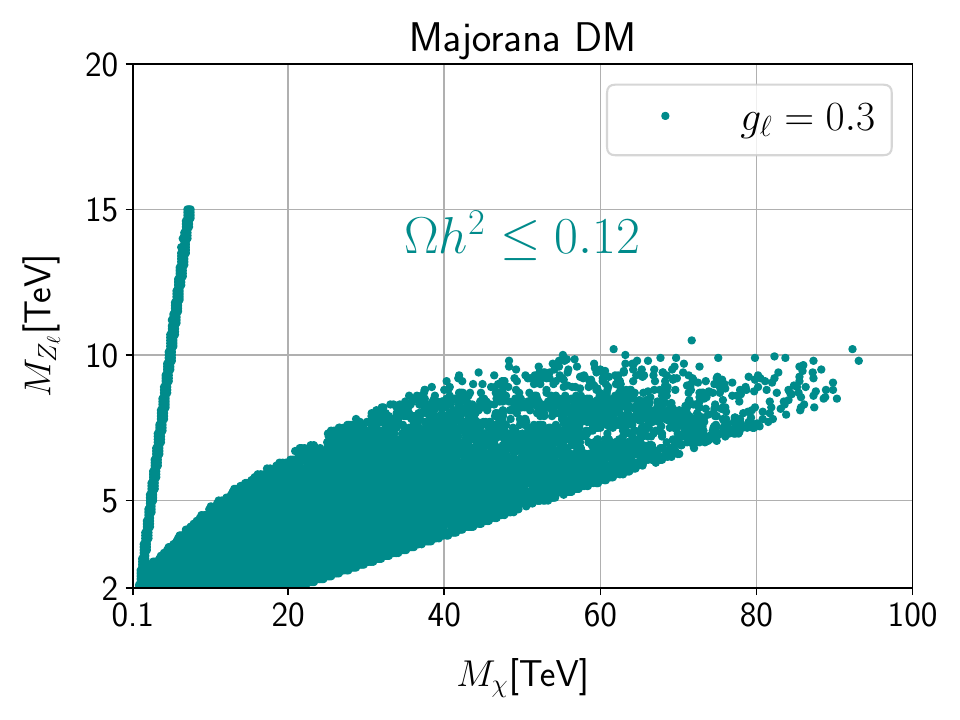}
\caption{Allowed parameter space by the relic density constraints when $g_\ell=0.3$.}
\label{all03}
\end{figure}
We show in Fig.~\ref{all08} the viable parameter when all the dark matter annihilation channels are included. As we expect, there are two main allowed regions, the resonance region which is not very generic and the other region where the gauge boson mass is always below $30$ TeV. Even in the resonance region, the gauge boson mass is below approximately $50$ TeV. It is remarkable that the upper bound on the symmetry breaking is just $30$ TeV in the most generic region. Therefore, one can be optimistic about the possibility to test this theory at colliders. We have used the value for the gauge coupling, $g_\ell=0.8$, which is basically the upper bound coming from perturbative to show the most conservative values for the upper bound on the gauge boson mass allowed by the cosmological bounds. However, the gauge coupling can be smaller and the upper bound could be even lower. 

In Fig.~\ref{all03} we show again the allowed parameter space but in this case $g_\ell=0.3$.
As one can see, the most generic region, outside the $Z_\ell$-resonance, tell us that the upper bound on the gauge boson mass is always below $10$ TeV. The most important result in this section is that as in the Dirac case studied in Ref.~\cite{Debnath:2023akj}, the cosmological bounds on the dark matter relic density implies an upper bound on the symmetry breaking for the spontaneous breaking of total lepton number in the multi-TeV region and one can hope to test the mechanism for Majorana neutrinos in the near future. 

In the Dirac case studied in Ref.~\cite{Debnath:2023akj} the upper bound on the symmetry breaking scale is larger than in the Majorana case because in the Majorana case studied in this article we have many annihilation channels that are velocity suppressed. In particular, the annihilation channels mediated by the new gauge boson, such as $\chi \chi \to Z_\ell \to \nu_i \nu_i, e_i^+ e_i^-$, are velocity suppressed and one cannot satisfy the cosmological bounds on the relic density when the symmetry breaking scale is above $50$ TeV. In the Dirac case studied in Ref.~\cite{Debnath:2023akj} the upper bound on the symmetry breaking scale is approximately $100$ TeV.   

In this theory, the dark matter-quark interactions are mediated only by the Higgses because the $Z_{\ell}$ gauge boson couples only to leptons at tree level. Here we are neglecting the kinetic mixing between the hypercharge and leptophilic gauge boson. The dark matter-nucleon spin-independent cross-section can be written as
\begin{figure}[t]
\centering
\includegraphics[width=0.45\textwidth]{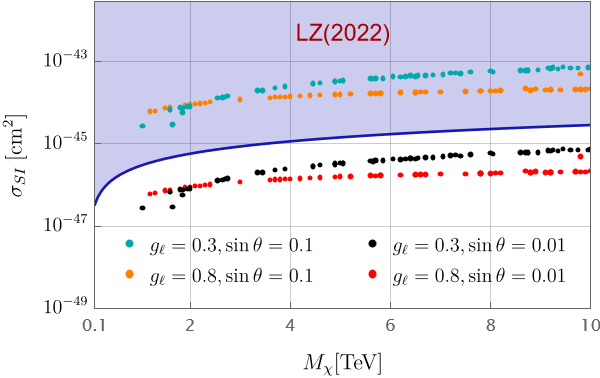}
\caption{Spin-independent dark matter-nucleon cross section.}
\label{DMDD}
\end{figure}

\begin{eqnarray}
    \sigma_{SI}=\frac{9 \sqrt{2}}{\pi}\frac{G_F}{M_h^4}\frac{g_{\ell}^2M_{\chi}^2\sin^2{\theta}}{M_{Z_{\ell}}^2\sin^2{\beta}}\frac{M_{\chi}^2 M_n^4}{(M_{\chi}+M_{n})^2} f_n^2, 
\end{eqnarray}
where $\sin{\theta}= U_{21}U_{11}$. In the above equation, $ M_n$ is the nucleon mass, and $ f_n=0.3$ is the effective Higgs-nucleon-nucleon effective coupling~\cite{Hoferichter:2017olk}. Notice that the cross section is suppressed by the mixing angle, $\theta$, which is the mixing angle between the Standard Model Higgs and the new Higgses. The mixing angle is naturally small when the new Higgses are heavy. 
Notice that if we include the kinetic mixing between the two Abelian gauge groups the new neutral gauge boson can also mediate this scattering process. This contribution to the cross section for the dark mater-nucleon scattering is highly suppressed by the mass of the new gauge boson but it is independent of the mixing angle between the Higgses. In this theory, the new gauge boson $Z_\ell$ mediates the dark matter-electron scattering but it is very suppressed by the gauge boson mass.
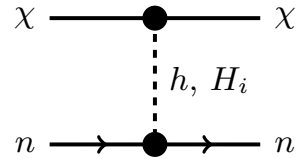
\begin{figure}[htbp]
\begin{tikzpicture}[line width=1.5 pt, node distance=1 cm and 1.5 cm, scale=1.40, transform shape]
\coordinate[label=left: $\chi$] (i1);
\coordinate[right=1cm of i1] (v1);
\coordinate[below=0.6cm of v1, label=right: {$h$, $H_i$}] (vaux);
\coordinate[right=1cm of v1, label=right:$\chi$] (f1);
\coordinate[below=1.2 cm of v1] (v2);
\coordinate[left=1 cm of v2, label=left: $n$] (i2);
\coordinate[right=1 cm of v2, label=right: $n$] (f2);
\draw[fermionnoarrow] (i1) -- (v1);
\draw[scalarnoarrow] (v1) -- (v2);
\draw[fermion] (i2) -- (v2); 
\draw[fermion] (v2) -- (f2); 
\draw[fermionnoarrow] (v1) -- (f1);
\draw[fill=black] (v1) circle (.1cm);
\draw[fill=black] (v2) circle (.1cm);
\end{tikzpicture}
\caption{Spin-independent nucleon-dark matter scattering mediated mainly by the Standard Model-like Higgs $h$, and the new Higgses $H_1$ and $H_2$.}
\label{fig:dirdetection}
\end{figure}

In Fig.~\ref{DMDD} we show the prediction for the dark matter-nucleon cross section for different values of the gauge coupling and the mixing angle $\theta$. Here we use $\sin \beta=0.54$ for illustration. As one can see, one can satisfy the experimental bounds when the mixing $\sin{\theta} \leq 0.01$. Notice that in this theory, one can have a viable dark matter candidate with mass even very close to the electroweak scale. 

One can consider different signatures for indirect detection experiments. For example, one can have the annihilation to charged leptons, $\chi \chi \to e_i^+ e_i^-$, but since these annihilation channels are velocity suppressed the theoretical predictions are very far from the current and future experimental bounds. One can consider also gamma line signatures but in this theory these cross sections mediated by the new gauge boson are also suppressed. For example, $\chi \chi \to \gamma \gamma$ mediated by the $Z_\ell$ is very suppressed by the large $Z_\ell$ gauge boson mass. We will study these signatures in detail in a future publication.
\section{$J$ DECAYS AND INTERACTIONS}
\label{Sec3}
\begin{figure}[h]
\centering
\includegraphics[width=0.45\textwidth]{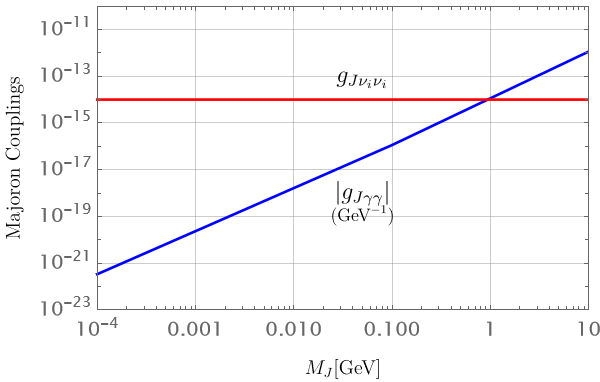}
\caption{Majoron couplings to neutrinos (in red) and to photons (in blue). Here we use $M_{Z_{\ell}}/g_{\ell}=10$ TeV, $M_{\psi}=2.5$ TeV and $M_{\eta}= 2$ TeV and $m_{\nu_i}=0.05$ eV }
\label{Majoron-couplins}
\end{figure}
\begin{figure}[h]
\centering
\includegraphics[width=0.45\textwidth]{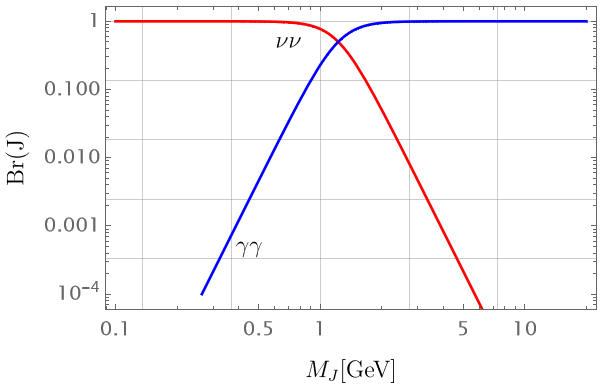}
\caption{Branching ratios for the Majoron decays vs. the Majoron mass when $M_{\psi}=2.5$ TeV and $M_{\eta}= 2$ TeV and $\sum_{i=1}^3 m_{\nu_i}=0.1$ eV.}
\label{BrJ}
\end{figure}
\begin{figure}[h]
\centering
\includegraphics[width=0.45\textwidth]{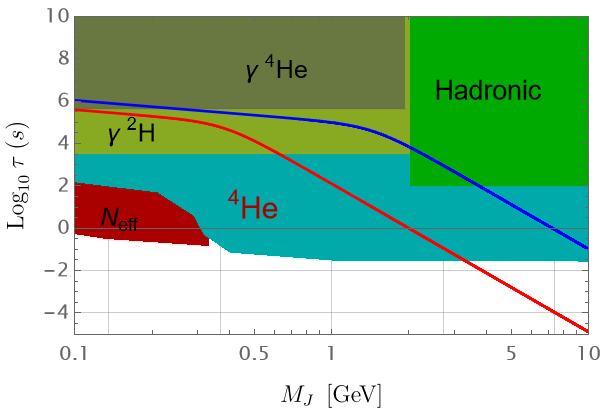}
\caption{Majoron lifetime vs. mass. For illustration we use $M_{Z_{\ell}}/g_{\ell}=10$ TeV, $M_{\psi}=2.5$ TeV, $M_{\eta}= 2$  TeV and $\sum_{i=1}^3 m_{\nu_i}=0.1$ eV for the blue line and $M_{Z_{\ell}}/g_{\ell}=6$ TeV, $M_{\psi}=300$ GeV, $M_{\eta}= 250$  GeV and $\sum_{i=1}^3 m_{\nu_i}=0.1$ eV for the red line. Here the color shaded regions are excluded by different bounds discussed in the text.}
\label{halfJ}
\end{figure}

In this theory, the Majoron mass is given by
\begin{equation}
M_J^2= \frac{\lambda_M}{4 \sqrt{2} \Lambda} \frac{M_{Z_\ell}^3}{g_\ell^3} \cos \beta.    
\end{equation}
Notice that $M_J \sim 10^{-4}$ GeV when $\lambda_M \sim 1$, $M_{Z_\ell}/g_\ell \sim 10$ TeV, $\cos \beta \sim 1$ and $\Lambda \sim M_{Pl}$, with $M_{Pl}$ being the Planck scale. Of course, the cutoff $\Lambda$ can be much smaller than the Planck scale and the Majoron could be much heavier. For a detailed discussion of the Majoron properties see for example Ref.~\cite{Heeck:2019guh}.

The effective coupling of the Majoron to photons is generated at one-loop level where inside the loop we have the new electrically charged fermions needed for anomaly cancellation. See Table.~\ref{anomalons} for the new fields needed for the anomaly cancellation. The effective $J (p)-\gamma (p_1)-\gamma (p_2)$ coupling can be written as
\begin{equation}
g_{J\gamma\gamma}= \sum_f \frac{c_f \ e_f^2}{\pi^2}
    \left( 1  +  2 m_f^2 \ C_0(s,m_f) \right).   
\label{gJgammagamma}    
\end{equation}
Here $c_f$ is the coupling between $J$ and the charged fermion with electric charge $e_f$ and mass $m_f$ inside the loop. In this case $s=(p_1 + p_2)^2=p^2$, where $p_1$ and $p_2$ are the momenta associated to the photons and $p$ is the Majoron momentum. In this model, $c_\eta=3 g_\ell/2 M_{Z_\ell}$, while $c_\Psi=-3 g_\ell/2 M_{Z_\ell}$. See Appendix for more details.
The loop function $C_0(s,m)$ can be written as
\begin{equation}
C_0 (s,m)=\frac{1}{2s} ln^2 \left( \frac{2 m^2 -s + \sqrt{s (s - 4 m^2)}}{ 2 m^2}\right).
\label{loop}
\end{equation}
We have used Package-X~\cite{Patel:2015tea} to perform the one-loop calculation. The coupling between the Majoron and (on-shell) neutrinos reads as
\begin{equation}
g_{J\nu_i\nu_i}= \frac{ 2 \ g_\ell  \ m_{\nu_i}}{M_{Z_\ell}}.  
\label{gJnunu}
\end{equation}
One can set strong bounds on the Majoron couplings to neutrinos using the constraints from Big Bang Nucleosynthesis, see Refs.~\cite{Escudero:2019gvw,Sandner:2023ptm}, when the Majoron is very light. As we will discuss below, the bounds on the couplings of the Majoron to photons are stronger, and one can rule out a large fraction of the parameter space.

Using the Majoron couplings in Eqs.(\ref{gJgammagamma}) and (\ref{gJnunu}) one can predict the decays of the Majoron.  In Fig.~\ref{Majoron-couplins} we show the numerical values for the Majoron couplings.
We show the value for the coupling to neutrinos when 
$m_{\nu_i}=0.05$ eV, while the coupling to two photons is very small for most of the Majoron mass values. Only, when the Majoron mass is above $1$ GeV, the couplings to photons is larger. Of course, the exact value of the coupling to photons depend of the new charged fermion masses. Notice that here we show the numerical values of the Majoron couplings to neutrinos in the case where the neutrino spectrum is quasi-degenerate. Since the lifetime of the Majoron does not change too much if we assume the Normal Hierarchy or Inverted Hierarchy, we show the couplings to neutrinos only in the quasi-degenerate case. 

In Fig.~\ref{BrJ} we show the branching ratios for the Majoron decays using $M_{Z_{\ell}}/g_{\ell}=10$ TeV, $M_{\psi}=2.5$ TeV and $M_{\eta}= 2$ TeV and $\sum_{i=1}^3 m_{\nu_i}=0.1$ eV. One can see that the decays to neutrinos dominate when the Majoron is light, while the branching ratio to photons is larger when the Majoron mass is above $1$ GeV. 

In Fig.~\ref{halfJ} we show the values for the lifetime of the Majoron. In most of the parameter space, the Majoron decays after Big Bang Nucleosynthesis. Therefore, the cosmological bounds are very important to set bounds on the Majoron mass and couplings to the Standard Model fields. In Fig.~\ref{halfJ} we show the bounds from Big Bang Nucleosynthesis as discussed in Ref.~\cite{Cadamuro_2012}.
The region in red is excluded by the $N_{eff}$ constraints, the region in cyan is excluded by
Helium overproduction, the region in yellow is excluded by the Deuterium photodissociation, the region in gray is excluded by Helium photodissociation, while the region in green is excluded by hadronic cascades. As one can appreciate, these bounds can exclude a large fraction of the allowed values for the Majoron lifetime and mass. We show the predictions for the Majoron lifetime in two cases: the blue line shows the prediction when we use $M_{Z_{\ell}}/g_{\ell}=10$ TeV, $M_{\psi}=2.5$ TeV, $M_{\eta}= 2$  TeV and $\sum_{i=1}^3 m_{\nu_i}=0.1$ eV and the red line corresponds to the case when $M_{Z_{\ell}}/g_{\ell}=6$ TeV, $M_{\psi}=300$ GeV, $M_{\eta}= 250$  GeV and $\sum_{i=1}^3 m_{\nu_i}=0.1$ eV. The first case is basically ruled out when the Majoron mass is below $10$ GeV, while in the second case one can satisfy the bounds when the Majoron mass is above $3$ GeV. 
%
\section{SUMMARY}
\label{Sec4}
We have discussed a simple theory for Majorana neutrinos where the total lepton number is a local gauge symmetry spontaneously broken below the multi-TeV scale. Since the total lepton number is not anomaly-free in the Standard Model, one needs to add new fermions to define an anomaly-free gauge theory. This type of theories predict a cold dark matter from anomaly cancellation with specific properties related to the origin of neutrinos masses. 

We discussed the case where the dark matter is also a Majorana fermion and studied in detail all annihilation channels to predict the region of the parameter space where one can obtain the correct relic density in agreement with all experimental constraints. This theory predicts a pseudo-Nambu-Goldstone boson that modify the annihilation channels in such way one can find the correct relic density when the dark matter is not very heavy. Since the dark matter is a Majorana fermion most of the annihilation channels through the new gauge boson are velocity suppressed and as a consequence the upper bound on the gauge boson mass is not very large. In the most generic region of the parameter space (outside the $Z_l$-resonance) one predicts that the upper bound on the symmetry breaking scale is below $30$ TeV. Therefore, one can hope to test this theory for neutrino masses at current or future collider experiments.

We have discussed the direct detection constraints and since the dark matter Majorana fermion interacts mainly to charged leptons through the new gauge boson, the experimental bounds allows us to have scenarios where the dark matter mass is very close to the electroweak scale when the mixing between the Standard Model Higgs and the new Higgses is small. 

We discussed the predictions for the Majoron couplings to neutrinos and photons. The Majoron in this theory is very long-lived but it cannot be a dark matter candidate.
The Majoron couplings to photons is generated at one-loop level where inside the loop one has the new charged fermions needed for anomaly cancellation. We discussed the predictions for the $J$ couplings and the bounds from Big Bang Nucleosynthesis, showing that these bounds rule out a large fraction of the parameter space for the Majoron mass and its lifetime.

{\small {\textit{Acknowledgments:}}
The work of H.D., P.F.P. and K.G.Q. is supported by the U.S. Department of Energy, Office of Science, Office of High Energy Physics, under Award Number DE-SC0024160.
This work made use of the High Performance Computing Resource in the Core Facility for Advanced Research Computing at Case Western Reserve University. }

\appendix
\begin{widetext}
\section{FEYNMAN RULES AND DECAYS}
\label{AppendixA}
Here we list all interactions used in this article:
\begin{align}
    \chi \chi Z_\ell  &: \hspace{0.5cm}  i \frac{3}{2}g_\ell \gamma^\mu \gamma^5, \\
    \nu \nu Z_\mu^\ell &: \hspace{0.5cm} -ig_\ell\gamma^\mu \gamma^5, \\
     N N Z_\mu^\ell &: \hspace{0.5cm} ig_\ell\gamma^\mu \gamma^5, \\  
     \Bar{e} e Z_\mu^\ell &: \hspace{0.5cm} i g_\ell \gamma^\mu, \\
     \chi \chi H_i &:\hspace{0.5cm} i \frac{3M_\chi g_\ell}{M_{Z_\ell}}\frac{U_{2i}}{\sin\beta}, \\
     H_i N N &: \hspace{0.5cm} i\frac{2g_\ell M_N}{M_{Z_\ell}\cos\beta}U_{3i},\\
     H_i Z_\ell^\mu Z_\ell^\nu &: \hspace{0.5cm} -i2g_\ell M_{Z_\ell}(2 U_{3i}\cos\beta+3U_{2i}\sin\beta)g_{\mu \nu}.
\end{align}
Notice that we are working in the basis where the $J$-interactions are invariant under the shift symmetry. We redefine the fields as follows:
\begin{eqnarray}
\nu_R &\to & e^{-i \sigma_\phi/2v_\phi} \nu_R, \\
e_R &\to & e^{-i \sigma_\phi/2v_\phi} e_R, \\
\ell_L &\to & e^{-i \sigma_\phi/2v_\phi} \ell_L,\\
\chi_L &\to & e^{i \sigma_s/2v_s} \chi_L,\\
\chi_R &\to & e^{-i \sigma_s/2v_s} \chi_R.
\end{eqnarray}
The Majoron coupling to photons reads as
\begin{eqnarray}
J (p) A^\mu (p_1, \lambda_1) A^\nu (p_2, \lambda_2)&:& - i g_{J\gamma\gamma} \ 
\epsilon^{\mu \nu \alpha \beta} p_{1 \alpha}  p_{2 \beta}. 
    \label{Jgammagamma}
\end{eqnarray}
The couplings of the Majoron with the neutral fields are given by 
\begin{align}
    \chi (p_1,s_1) \chi (p_2,s_2) J (p) &: \hspace{0.5cm} i\frac{3g_\ell}{2M_{Z_\ell}}  p_\mu \gamma^\mu \gamma^5,\\
    Z_\mu^\ell (p_1,\lambda_1) H_i (p_2) J (p) &: \hspace{0.5cm} 2 i p_\mu \left( 2 g_\ell \cos\beta U_{3i} + 3 g_\ell \sin\beta U_{2i} \right), \\
    N (p_1,s_1) N (p_2, s_2) J (p) &: \hspace{0.5cm} -i\frac{g_\ell}{M_{Z_\ell}}p_\mu \gamma^\mu \gamma^5, \\
    \nu_i (p_1,s_1) \nu_i (p_2, s_2) J (p) &: \hspace{0.5cm} i\frac{g_\ell}{M_{Z_\ell}}p_\mu \gamma^\mu \gamma^5.
    \label{nunuJ}
\end{align}
In order to compute the Majoron coupling to photons we need to use a specific model for anomaly cancellation. The Yukawa interactions for the fields needed for anomaly cancellation listed in Table~\ref{anomalons} are given by 
\begin{eqnarray}
-\mathcal{L}_Y^L &=&  y_1 \bar{\Psi}_LH\eta_R + y_2 \bar{\Psi}_R H\eta_L +  y_3 \bar{\Psi}_L\tilde{H}\chi_R+y_4\bar{\Psi}_R\tilde{H}\chi_L  \nonumber \\
&+& y_{\Psi} \bar{\Psi}_L\Psi_R S^*
+ y_{\eta}\bar{\eta}_R \eta_L S^* + y_{\chi}\bar{\chi}_R \chi_L S^*  + 
\lambda_\chi \chi^T_L C \chi_L S^*  + \tilde{\lambda}_\chi \chi^T_R C \chi_R S 
 +  {\rm{h.c.}}.
    \label{eq:LB}
\end{eqnarray}

In the simple theory discussed in Ref.~\cite{Debnath:2023akj} the physical fields, $\eta^-=\eta_L^- + \eta_R^-$ and $\Psi^-=\Psi^-_L + \Psi^-_R$, couple to the Majoron in the following way:
\begin{align}
J (p) \overline{\eta^-} (p_1,s_1) \eta^- (p_2,s_2)&: \frac{3 i g_\ell}{2 M_{Z_\ell}} p_\mu \gamma^\mu \gamma^5, \\
J (p) \overline{\Psi^-} (p_1,s_1) \Psi^- (p_2,s_2)&:
-\frac{3 i g_\ell}{2 M_{Z_\ell}} p_\mu \gamma^\mu \gamma^5.
\end{align}
We work in the basis where we redefine the extra fields as follows:
\begin{eqnarray}
\eta_R &\to & e^{-i \sigma_s/2v_s} \eta_R, \\
\eta_L &\to & e^{i \sigma_s/2v_s} \eta_L,\\
\Psi_L &\to & e^{-i \sigma_s/2v_s} \Psi_L,\\
\Psi_R &\to & e^{i \sigma_s/2v_s} \Psi_R.
\end{eqnarray}
\begin{table}[t]
\begin{tabular}{||c c c c c||} 
 \hline
 \hline
 Fields & $SU(3)_C$ & $SU(2)_L$ &$ U(1)_Y$  &$U(1)_\ell$\\ [1.5 ex]
 \hline\hline
 $\Psi_L = \begin{pmatrix}
 \Psi_L^0\\
 \Psi_L^-
 \end{pmatrix}$ &$ \mathbf{1} $& $ \mathbf{2} $ & $-\frac{1}{2}$ & $\ell_1$  \\ [2ex]
 \hline
 $\Psi_R = \begin{pmatrix}
 \Psi_R^0\\
 \Psi_R^-
 \end{pmatrix}$ & $ \mathbf{1} $ & $\mathbf{2}$ & $-\frac{1}{2}$ & $\ell_2$ \\
 \hline
 $\eta_R$ & $ \mathbf{1} $ & $ \mathbf{1} $ & $-1$ & $\ell_1$ \\[2ex]
 \hline
 $\eta_L$ & $ \mathbf{1} $ & $ \mathbf{1} $ & $-1$ & $\ell_2$ \\[2ex]
 \hline
 $\chi_R$ & $ \mathbf{1} $ & $ \mathbf{1} $ & 0 & $\ell_1$\\  [2ex]
 \hline
 $\chi_L$& $\mathbf{1} $ & $\mathbf{1} $ & 0 & $\ell_2$\\[2 ex]
 \hline
\end{tabular}
\caption{\label{1} Fermions needed for anomaly cancellation with $\ell_1 - \ell_2  = - 3$~\cite{Duerr:2013dza}. In the Majorana case $\ell_1=-\ell_2=-3/2$.}
\label{anomalons}
\end{table}
The decay widths for the Majoron are given by
\begin{eqnarray}
\Gamma (J \to \nu \nu) &=& \sum_{i=1}^3 \Gamma (J \to \nu_i \nu_i)=\frac{M_J \ g_\ell^2}{4 \pi M_{Z_\ell}^2} \sum_{i=1}^3 m_{\nu_i}^2 \ \sqrt{1-\frac{4 m_{\nu_i}^2}{M^2_J}}, \\
\Gamma (J \to \gamma \gamma)&=&\frac{9 \alpha^2 g_{\ell}^2}{4 \pi^3 M_{Z_{\ell}}^2} M_J^3\left|M_{\eta}^2 C_0(M_J^2,M_{\eta})-M_{\psi}^2 C_0(M_J^2,M_{\psi})\right|^2. 
\end{eqnarray}
\section{BRANCHING RATIOS FOR THE DM ANNIHILATION CHANNELS}
\label{AppendixB}
\begin{figure}[h]
\centering
\includegraphics[width=0.70\textwidth]{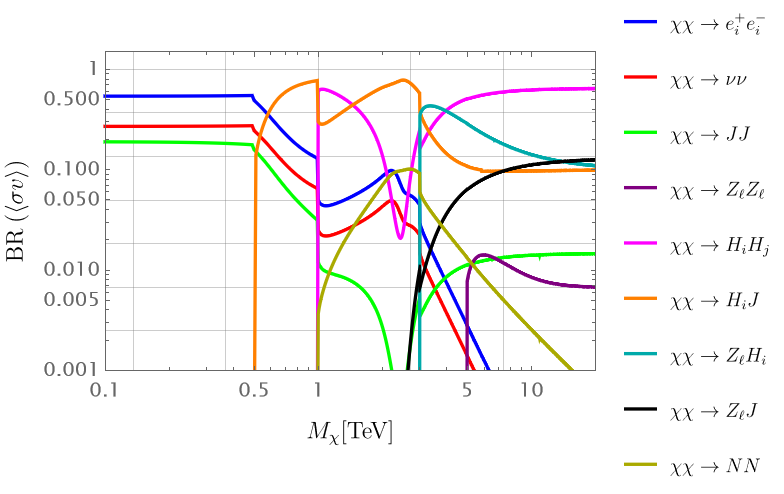}
        \caption{Branching ratio of the thermal average cross-section for the dominant annihilation channels as a function of dark matter mass. Here we use $M_{Z_\ell}= 5$ TeV, $g_\ell = 0.8$, $M_N = M_{H_i} = 1 $ TeV and the freeze-out temperature, $x_f=25$.
}
        \label{BR1}
\end{figure}
\begin{figure}[h]
\centering
\includegraphics[width=0.70\textwidth]{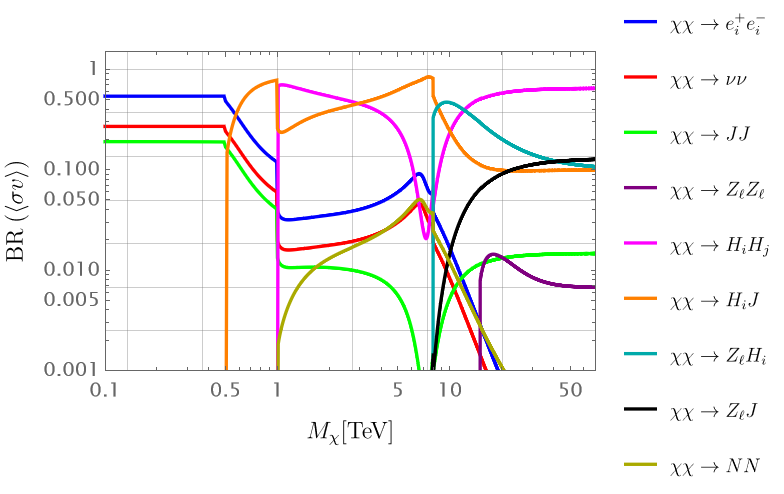}
        \caption{Branching ratio of the thermal average cross-section for the dominant annihilation channels as a function of dark matter mass. Here we used $M_{Z_\ell}= 15$ TeV, $g_\ell = 0.8$, $M_N = M_{H_i} = 1 $ TeV and the freeze-out temperature, $x_f=25$.
}
        \label{BR2}
\end{figure}
\begin{figure}[h]
\centering
\includegraphics[width=0.70\textwidth]{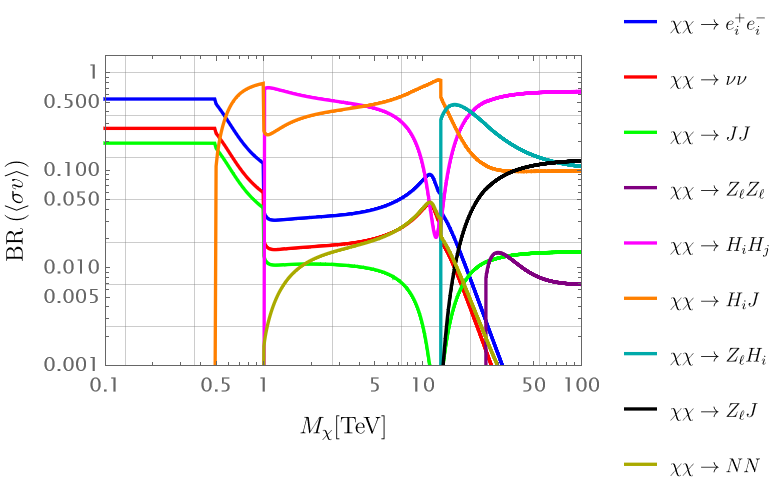}
        \caption{Branching ratio of the thermal average cross-section for the dominant annihilation channels as a function of dark matter mass. Here we use $M_{Z_\ell}= 25$ TeV, $g_\ell = 0.8$, $M_N = M_{H_i} = 1 $ TeV and the freeze-out temperature, $x_f=25$.
}
        \label{BR3}
\end{figure}

In order to appreciate the contribution of the different dark matter annihilation channels we show here the branching ratios for the different channels in some representative scenarios.
In Fig.~\ref{BR1} we show the branching ratios when $M_{Z_\ell}= 5$ TeV, $g_\ell = 0.8$, $M_N = M_{H_i} = 1 $ TeV and the freeze-out temperature, $x_f=25$. In Figs.~\ref{BR2} and~\ref{BR3} we show the branching ratios when $M_{Z_\ell}= 15$ TeV and $M_{Z_\ell}= 25$ TeV, respectively. We use the same values for the other parameters as in Fig.~\ref{BR1}.
Notice that in all these scenarios the annihilation into charged leptons dominate at low DM mass, while at large values of the DM mass the annihilation into the new Higgses is the most important annihilation channel.
\end{widetext}
\newpage
\bibliography{refs}

\begin{thebibliography}{25}%
\makeatletter
\providecommand \@ifxundefined [1]{%
 \@ifx{#1\undefined}
}%
\providecommand \@ifnum [1]{%
 \ifnum #1\expandafter \@firstoftwo
 \else \expandafter \@secondoftwo
 \fi
}%
\providecommand \@ifx [1]{%
 \ifx #1\expandafter \@firstoftwo
 \else \expandafter \@secondoftwo
 \fi
}%
\providecommand \natexlab [1]{#1}%
\providecommand \enquote  [1]{``#1''}%
\providecommand \bibnamefont  [1]{#1}%
\providecommand \bibfnamefont [1]{#1}%
\providecommand \citenamefont [1]{#1}%
\providecommand \href@noop [0]{\@secondoftwo}%
\providecommand \href [0]{\begingroup \@sanitize@url \@href}%
\providecommand \@href[1]{\@@startlink{#1}\@@href}%
\providecommand \@@href[1]{\endgroup#1\@@endlink}%
\providecommand \@sanitize@url [0]{\catcode `\\12\catcode `\$12\catcode `\&12\catcode `\#12\catcode `\^12\catcode `\_12\catcode `\%12\relax}%
\providecommand \@@startlink[1]{}%
\providecommand \@@endlink[0]{}%
\providecommand \url  [0]{\begingroup\@sanitize@url \@url }%
\providecommand \@url [1]{\endgroup\@href {#1}{\urlprefix }}%
\providecommand \urlprefix  [0]{URL }%
\providecommand \Eprint [0]{\href }%
\providecommand \doibase [0]{http://dx.doi.org/}%
\providecommand \selectlanguage [0]{\@gobble}%
\providecommand \bibinfo  [0]{\@secondoftwo}%
\providecommand \bibfield  [0]{\@secondoftwo}%
\providecommand \translation [1]{[#1]}%
\providecommand \BibitemOpen [0]{}%
\providecommand \bibitemStop [0]{}%
\providecommand \bibitemNoStop [0]{.\EOS\space}%
\providecommand \EOS [0]{\spacefactor3000\relax}%
\providecommand \BibitemShut  [1]{\csname bibitem#1\endcsname}%
\let\auto@bib@innerbib\@empty
\bibitem [{\citenamefont {Minkowski}(1977)}]{Minkowski:1977sc}%
  \BibitemOpen
  \bibfield  {author} {\bibinfo {author} {\bibfnamefont {P.}~\bibnamefont {Minkowski}},\ }\href {\doibase 10.1016/0370-2693(77)90435-X} {\bibfield  {journal} {\bibinfo  {journal} {Phys. Lett. B}\ }\textbf {\bibinfo {volume} {67}},\ \bibinfo {pages} {421} (\bibinfo {year} {1977})}\BibitemShut {NoStop}%
\bibitem [{\citenamefont {Gell-Mann}\ \emph {et~al.}(1979)\citenamefont {Gell-Mann}, \citenamefont {Ramond},\ and\ \citenamefont {Slansky}}]{Gell-Mann:1979vob}%
  \BibitemOpen
  \bibfield  {author} {\bibinfo {author} {\bibfnamefont {M.}~\bibnamefont {Gell-Mann}}, \bibinfo {author} {\bibfnamefont {P.}~\bibnamefont {Ramond}}, \ and\ \bibinfo {author} {\bibfnamefont {R.}~\bibnamefont {Slansky}},\ }\href@noop {} {\bibfield  {journal} {\bibinfo  {journal} {Conf. Proc. C}\ }\textbf {\bibinfo {volume} {790927}},\ \bibinfo {pages} {315} (\bibinfo {year} {1979})},\ \Eprint {http://arxiv.org/abs/1306.4669} {arXiv:1306.4669 [hep-th]} \BibitemShut {NoStop}%
\bibitem [{\citenamefont {Yanagida}(1979)}]{Yanagida:1979as}%
  \BibitemOpen
  \bibfield  {author} {\bibinfo {author} {\bibfnamefont {T.}~\bibnamefont {Yanagida}},\ }\href@noop {} {\bibfield  {journal} {\bibinfo  {journal} {Conf. Proc. C}\ }\textbf {\bibinfo {volume} {7902131}},\ \bibinfo {pages} {95} (\bibinfo {year} {1979})}\BibitemShut {NoStop}%
\bibitem [{\citenamefont {Mohapatra}\ and\ \citenamefont {Senjanovic}(1980)}]{Mohapatra:1979ia}%
  \BibitemOpen
  \bibfield  {author} {\bibinfo {author} {\bibfnamefont {R.~N.}\ \bibnamefont {Mohapatra}}\ and\ \bibinfo {author} {\bibfnamefont {G.}~\bibnamefont {Senjanovic}},\ }\href {\doibase 10.1103/PhysRevLett.44.912} {\bibfield  {journal} {\bibinfo  {journal} {Phys. Rev. Lett.}\ }\textbf {\bibinfo {volume} {44}},\ \bibinfo {pages} {912} (\bibinfo {year} {1980})}\BibitemShut {NoStop}%
\bibitem [{\citenamefont {Fileviez~Perez}\ and\ \citenamefont {Wise}(2011)}]{FileviezPerez:2011pt}%
  \BibitemOpen
  \bibfield  {author} {\bibinfo {author} {\bibfnamefont {P.}~\bibnamefont {Fileviez~Perez}}\ and\ \bibinfo {author} {\bibfnamefont {M.~B.}\ \bibnamefont {Wise}},\ }\href {\doibase 10.1007/JHEP08(2011)068} {\bibfield  {journal} {\bibinfo  {journal} {JHEP}\ }\textbf {\bibinfo {volume} {08}},\ \bibinfo {pages} {068} (\bibinfo {year} {2011})},\ \Eprint {http://arxiv.org/abs/1106.0343} {arXiv:1106.0343 [hep-ph]} \BibitemShut {NoStop}%
\bibitem [{\citenamefont {Duerr}\ \emph {et~al.}(2013)\citenamefont {Duerr}, \citenamefont {Fileviez~Perez},\ and\ \citenamefont {Wise}}]{Duerr:2013dza}%
  \BibitemOpen
  \bibfield  {author} {\bibinfo {author} {\bibfnamefont {M.}~\bibnamefont {Duerr}}, \bibinfo {author} {\bibfnamefont {P.}~\bibnamefont {Fileviez~Perez}}, \ and\ \bibinfo {author} {\bibfnamefont {M.~B.}\ \bibnamefont {Wise}},\ }\href {\doibase 10.1103/PhysRevLett.110.231801} {\bibfield  {journal} {\bibinfo  {journal} {Phys. Rev. Lett.}\ }\textbf {\bibinfo {volume} {110}},\ \bibinfo {pages} {231801} (\bibinfo {year} {2013})},\ \Eprint {http://arxiv.org/abs/1304.0576} {arXiv:1304.0576 [hep-ph]} \BibitemShut {NoStop}%
\bibitem [{\citenamefont {Fileviez~Perez}\ \emph {et~al.}(2014)\citenamefont {Fileviez~Perez}, \citenamefont {Ohmer},\ and\ \citenamefont {Patel}}]{FileviezPerez:2014lnj}%
  \BibitemOpen
  \bibfield  {author} {\bibinfo {author} {\bibfnamefont {P.}~\bibnamefont {Fileviez~Perez}}, \bibinfo {author} {\bibfnamefont {S.}~\bibnamefont {Ohmer}}, \ and\ \bibinfo {author} {\bibfnamefont {H.~H.}\ \bibnamefont {Patel}},\ }\href {\doibase 10.1016/j.physletb.2014.06.057} {\bibfield  {journal} {\bibinfo  {journal} {Phys. Lett. B}\ }\textbf {\bibinfo {volume} {735}},\ \bibinfo {pages} {283} (\bibinfo {year} {2014})},\ \Eprint {http://arxiv.org/abs/1403.8029} {arXiv:1403.8029 [hep-ph]} \BibitemShut {NoStop}%
\bibitem [{\citenamefont {Foot}\ \emph {et~al.}(1989)\citenamefont {Foot}, \citenamefont {Joshi},\ and\ \citenamefont {Lew}}]{Foot:1989ts}%
  \BibitemOpen
  \bibfield  {author} {\bibinfo {author} {\bibfnamefont {R.}~\bibnamefont {Foot}}, \bibinfo {author} {\bibfnamefont {G.~C.}\ \bibnamefont {Joshi}}, \ and\ \bibinfo {author} {\bibfnamefont {H.}~\bibnamefont {Lew}},\ }\href {\doibase 10.1103/PhysRevD.40.2487} {\bibfield  {journal} {\bibinfo  {journal} {Phys. Rev. D}\ }\textbf {\bibinfo {volume} {40}},\ \bibinfo {pages} {2487} (\bibinfo {year} {1989})}\BibitemShut {NoStop}%
\bibitem [{\citenamefont {Debnath}\ and\ \citenamefont {Fileviez~Perez}(2023)}]{Debnath:2023akj}%
  \BibitemOpen
  \bibfield  {author} {\bibinfo {author} {\bibfnamefont {H.}~\bibnamefont {Debnath}}\ and\ \bibinfo {author} {\bibfnamefont {P.}~\bibnamefont {Fileviez~Perez}},\ }\href {\doibase 10.1103/PhysRevD.108.075009} {\bibfield  {journal} {\bibinfo  {journal} {Phys. Rev. D}\ }\textbf {\bibinfo {volume} {108}},\ \bibinfo {pages} {075009} (\bibinfo {year} {2023})},\ \Eprint {http://arxiv.org/abs/2307.03646} {arXiv:2307.03646 [hep-ph]} \BibitemShut {NoStop}%
\bibitem [{\citenamefont {Chikashige}\ \emph {et~al.}(1981)\citenamefont {Chikashige}, \citenamefont {Mohapatra},\ and\ \citenamefont {Peccei}}]{Chikashige:1980ui}%
  \BibitemOpen
  \bibfield  {author} {\bibinfo {author} {\bibfnamefont {Y.}~\bibnamefont {Chikashige}}, \bibinfo {author} {\bibfnamefont {R.~N.}\ \bibnamefont {Mohapatra}}, \ and\ \bibinfo {author} {\bibfnamefont {R.~D.}\ \bibnamefont {Peccei}},\ }\href {\doibase 10.1016/0370-2693(81)90011-3} {\bibfield  {journal} {\bibinfo  {journal} {Phys. Lett. B}\ }\textbf {\bibinfo {volume} {98}},\ \bibinfo {pages} {265} (\bibinfo {year} {1981})}\BibitemShut {NoStop}%
\bibitem [{\citenamefont {Schwaller}\ \emph {et~al.}(2013)\citenamefont {Schwaller}, \citenamefont {Tait},\ and\ \citenamefont {Vega-Morales}}]{Schwaller:2013hqa}%
  \BibitemOpen
  \bibfield  {author} {\bibinfo {author} {\bibfnamefont {P.}~\bibnamefont {Schwaller}}, \bibinfo {author} {\bibfnamefont {T.~M.~P.}\ \bibnamefont {Tait}}, \ and\ \bibinfo {author} {\bibfnamefont {R.}~\bibnamefont {Vega-Morales}},\ }\href {\doibase 10.1103/PhysRevD.88.035001} {\bibfield  {journal} {\bibinfo  {journal} {Phys. Rev. D}\ }\textbf {\bibinfo {volume} {88}},\ \bibinfo {pages} {035001} (\bibinfo {year} {2013})},\ \Eprint {http://arxiv.org/abs/1305.1108} {arXiv:1305.1108 [hep-ph]} \BibitemShut {NoStop}%
\bibitem [{\citenamefont {Fileviez~P\'erez}\ \emph {et~al.}(2019)\citenamefont {Fileviez~P\'erez}, \citenamefont {Murgui},\ and\ \citenamefont {Plascencia}}]{FileviezPerez:2019cyn}%
  \BibitemOpen
  \bibfield  {author} {\bibinfo {author} {\bibfnamefont {P.}~\bibnamefont {Fileviez~P\'erez}}, \bibinfo {author} {\bibfnamefont {C.}~\bibnamefont {Murgui}}, \ and\ \bibinfo {author} {\bibfnamefont {A.~D.}\ \bibnamefont {Plascencia}},\ }\href {\doibase 10.1103/PhysRevD.100.035041} {\bibfield  {journal} {\bibinfo  {journal} {Phys. Rev. D}\ }\textbf {\bibinfo {volume} {100}},\ \bibinfo {pages} {035041} (\bibinfo {year} {2019})},\ \Eprint {http://arxiv.org/abs/1905.06344} {arXiv:1905.06344 [hep-ph]} \BibitemShut {NoStop}%
\bibitem [{\citenamefont {Carena}\ \emph {et~al.}(2023)\citenamefont {Carena}, \citenamefont {Li}, \citenamefont {Ou},\ and\ \citenamefont {Wang}}]{Carena:2022qpf}%
  \BibitemOpen
  \bibfield  {author} {\bibinfo {author} {\bibfnamefont {M.}~\bibnamefont {Carena}}, \bibinfo {author} {\bibfnamefont {Y.-Y.}\ \bibnamefont {Li}}, \bibinfo {author} {\bibfnamefont {T.}~\bibnamefont {Ou}}, \ and\ \bibinfo {author} {\bibfnamefont {Y.}~\bibnamefont {Wang}},\ }\href {\doibase 10.1007/JHEP02(2023)139} {\bibfield  {journal} {\bibinfo  {journal} {JHEP}\ }\textbf {\bibinfo {volume} {02}},\ \bibinfo {pages} {139} (\bibinfo {year} {2023})},\ \Eprint {http://arxiv.org/abs/2210.14352} {arXiv:2210.14352 [hep-ph]} \BibitemShut {NoStop}%
\bibitem [{\citenamefont {Madge}\ and\ \citenamefont {Schwaller}(2019)}]{Madge:2018gfl}%
  \BibitemOpen
  \bibfield  {author} {\bibinfo {author} {\bibfnamefont {E.}~\bibnamefont {Madge}}\ and\ \bibinfo {author} {\bibfnamefont {P.}~\bibnamefont {Schwaller}},\ }\href {\doibase 10.1007/JHEP02(2019)048} {\bibfield  {journal} {\bibinfo  {journal} {JHEP}\ }\textbf {\bibinfo {volume} {02}},\ \bibinfo {pages} {048} (\bibinfo {year} {2019})},\ \Eprint {http://arxiv.org/abs/1809.09110} {arXiv:1809.09110 [hep-ph]} \BibitemShut {NoStop}%
\bibitem [{\citenamefont {Fileviez~Perez}(2015)}]{FileviezPerez:2015mlm}%
  \BibitemOpen
  \bibfield  {author} {\bibinfo {author} {\bibfnamefont {P.}~\bibnamefont {Fileviez~Perez}},\ }\href {\doibase 10.1016/j.physrep.2015.09.001} {\bibfield  {journal} {\bibinfo  {journal} {Phys. Rept.}\ }\textbf {\bibinfo {volume} {597}},\ \bibinfo {pages} {1} (\bibinfo {year} {2015})},\ \Eprint {http://arxiv.org/abs/1501.01886} {arXiv:1501.01886 [hep-ph]} \BibitemShut {NoStop}%
\bibitem [{\citenamefont {Aranda}\ \emph {et~al.}(2015)\citenamefont {Aranda}, \citenamefont {Jim\'enez},\ and\ \citenamefont {Vaquera-Araujo}}]{Aranda:2014zta}%
  \BibitemOpen
  \bibfield  {author} {\bibinfo {author} {\bibfnamefont {A.}~\bibnamefont {Aranda}}, \bibinfo {author} {\bibfnamefont {E.}~\bibnamefont {Jim\'enez}}, \ and\ \bibinfo {author} {\bibfnamefont {C.~A.}\ \bibnamefont {Vaquera-Araujo}},\ }\href {\doibase 10.1007/JHEP01(2015)070} {\bibfield  {journal} {\bibinfo  {journal} {JHEP}\ }\textbf {\bibinfo {volume} {01}},\ \bibinfo {pages} {070} (\bibinfo {year} {2015})},\ \Eprint {http://arxiv.org/abs/1410.7508} {arXiv:1410.7508 [hep-ph]} \BibitemShut {NoStop}%
\bibitem [{\citenamefont {Gondolo}\ and\ \citenamefont {Gelmini}(1991)}]{Gondolo:1990dk}%
  \BibitemOpen
  \bibfield  {author} {\bibinfo {author} {\bibfnamefont {P.}~\bibnamefont {Gondolo}}\ and\ \bibinfo {author} {\bibfnamefont {G.}~\bibnamefont {Gelmini}},\ }\href {\doibase 10.1016/0550-3213(91)90438-4} {\bibfield  {journal} {\bibinfo  {journal} {Nucl. Phys. B}\ }\textbf {\bibinfo {volume} {360}},\ \bibinfo {pages} {145} (\bibinfo {year} {1991})}\BibitemShut {NoStop}%
\bibitem [{\citenamefont {Aghanim}\ \emph {et~al.}(2020)\citenamefont {Aghanim} \emph {et~al.}}]{Planck:2018vyg}%
  \BibitemOpen
  \bibfield  {author} {\bibinfo {author} {\bibfnamefont {N.}~\bibnamefont {Aghanim}} \emph {et~al.} (\bibinfo {collaboration} {Planck}),\ }\href {\doibase 10.1051/0004-6361/201833910} {\bibfield  {journal} {\bibinfo  {journal} {Astron. Astrophys.}\ }\textbf {\bibinfo {volume} {641}},\ \bibinfo {pages} {A6} (\bibinfo {year} {2020})},\ \bibinfo {note} {[Erratum: Astron.Astrophys. 652, C4 (2021)]},\ \Eprint {http://arxiv.org/abs/1807.06209} {arXiv:1807.06209 [astro-ph.CO]} \BibitemShut {NoStop}%
\bibitem [{\citenamefont {Alioli}\ \emph {et~al.}(2018)\citenamefont {Alioli}, \citenamefont {Farina}, \citenamefont {Pappadopulo},\ and\ \citenamefont {Ruderman}}]{Alioli:2017nzr}%
  \BibitemOpen
  \bibfield  {author} {\bibinfo {author} {\bibfnamefont {S.}~\bibnamefont {Alioli}}, \bibinfo {author} {\bibfnamefont {M.}~\bibnamefont {Farina}}, \bibinfo {author} {\bibfnamefont {D.}~\bibnamefont {Pappadopulo}}, \ and\ \bibinfo {author} {\bibfnamefont {J.~T.}\ \bibnamefont {Ruderman}},\ }\href {\doibase 10.1103/PhysRevLett.120.101801} {\bibfield  {journal} {\bibinfo  {journal} {Phys. Rev. Lett.}\ }\textbf {\bibinfo {volume} {120}},\ \bibinfo {pages} {101801} (\bibinfo {year} {2018})},\ \Eprint {http://arxiv.org/abs/1712.02347} {arXiv:1712.02347 [hep-ph]} \BibitemShut {NoStop}%
\bibitem [{\citenamefont {Hoferichter}\ \emph {et~al.}(2017)\citenamefont {Hoferichter}, \citenamefont {Klos}, \citenamefont {Men\'endez},\ and\ \citenamefont {Schwenk}}]{Hoferichter:2017olk}%
  \BibitemOpen
  \bibfield  {author} {\bibinfo {author} {\bibfnamefont {M.}~\bibnamefont {Hoferichter}}, \bibinfo {author} {\bibfnamefont {P.}~\bibnamefont {Klos}}, \bibinfo {author} {\bibfnamefont {J.}~\bibnamefont {Men\'endez}}, \ and\ \bibinfo {author} {\bibfnamefont {A.}~\bibnamefont {Schwenk}},\ }\href {\doibase 10.1103/PhysRevLett.119.181803} {\bibfield  {journal} {\bibinfo  {journal} {Phys. Rev. Lett.}\ }\textbf {\bibinfo {volume} {119}},\ \bibinfo {pages} {181803} (\bibinfo {year} {2017})},\ \Eprint {http://arxiv.org/abs/1708.02245} {arXiv:1708.02245 [hep-ph]} \BibitemShut {NoStop}%
\bibitem [{\citenamefont {Heeck}\ and\ \citenamefont {Patel}(2019)}]{Heeck:2019guh}%
  \BibitemOpen
  \bibfield  {author} {\bibinfo {author} {\bibfnamefont {J.}~\bibnamefont {Heeck}}\ and\ \bibinfo {author} {\bibfnamefont {H.~H.}\ \bibnamefont {Patel}},\ }\href {\doibase 10.1103/PhysRevD.100.095015} {\bibfield  {journal} {\bibinfo  {journal} {Phys. Rev. D}\ }\textbf {\bibinfo {volume} {100}},\ \bibinfo {pages} {095015} (\bibinfo {year} {2019})},\ \Eprint {http://arxiv.org/abs/1909.02029} {arXiv:1909.02029 [hep-ph]} \BibitemShut {NoStop}%
\bibitem [{\citenamefont {Patel}(2015)}]{Patel:2015tea}%
  \BibitemOpen
  \bibfield  {author} {\bibinfo {author} {\bibfnamefont {H.~H.}\ \bibnamefont {Patel}},\ }\href {\doibase 10.1016/j.cpc.2015.08.017} {\bibfield  {journal} {\bibinfo  {journal} {Comput. Phys. Commun.}\ }\textbf {\bibinfo {volume} {197}},\ \bibinfo {pages} {276} (\bibinfo {year} {2015})},\ \Eprint {http://arxiv.org/abs/1503.01469} {arXiv:1503.01469 [hep-ph]} \BibitemShut {NoStop}%
\bibitem [{\citenamefont {Escudero}\ and\ \citenamefont {Witte}(2020)}]{Escudero:2019gvw}%
  \BibitemOpen
  \bibfield  {author} {\bibinfo {author} {\bibfnamefont {M.}~\bibnamefont {Escudero}}\ and\ \bibinfo {author} {\bibfnamefont {S.~J.}\ \bibnamefont {Witte}},\ }\href {\doibase 10.1140/epjc/s10052-020-7854-5} {\bibfield  {journal} {\bibinfo  {journal} {Eur. Phys. J. C}\ }\textbf {\bibinfo {volume} {80}},\ \bibinfo {pages} {294} (\bibinfo {year} {2020})},\ \Eprint {http://arxiv.org/abs/1909.04044} {arXiv:1909.04044 [astro-ph.CO]} \BibitemShut {NoStop}%
\bibitem [{\citenamefont {Sandner}\ \emph {et~al.}(2023)\citenamefont {Sandner}, \citenamefont {Escudero},\ and\ \citenamefont {Witte}}]{Sandner:2023ptm}%
  \BibitemOpen
  \bibfield  {author} {\bibinfo {author} {\bibfnamefont {S.}~\bibnamefont {Sandner}}, \bibinfo {author} {\bibfnamefont {M.}~\bibnamefont {Escudero}}, \ and\ \bibinfo {author} {\bibfnamefont {S.~J.}\ \bibnamefont {Witte}},\ }\href@noop {} {\  (\bibinfo {year} {2023})},\ \Eprint {http://arxiv.org/abs/2305.01692} {arXiv:2305.01692 [hep-ph]} \BibitemShut {NoStop}%
\bibitem [{\citenamefont {Cadamuro}\ and\ \citenamefont {Redondo}(2012)}]{Cadamuro_2012}%
  \BibitemOpen
  \bibfield  {author} {\bibinfo {author} {\bibfnamefont {D.}~\bibnamefont {Cadamuro}}\ and\ \bibinfo {author} {\bibfnamefont {J.}~\bibnamefont {Redondo}},\ }\href {\doibase 10.1088/1475-7516/2012/02/032} {\bibfield  {journal} {\bibinfo  {journal} {Journal of Cosmology and Astroparticle Physics}\ }\textbf {\bibinfo {volume} {2012}},\ \bibinfo {pages} {032–032} (\bibinfo {year} {2012})}\BibitemShut {NoStop}%
\end{thebibliography}%

\end{document}